\documentclass[prr, superscriptaddress, amsmath, amssymb, aps, twocolumn]{revtex4-2}
\usepackage[utf8]{inputenc}
\usepackage[english]{babel}
\usepackage{physics}
\usepackage{graphics}
\usepackage{graphicx}
\usepackage{float}
 \usepackage{color}
 \usepackage{soul}
 \usepackage[colorlinks=true,linkcolor=blue]{hyperref}

\usepackage{bm}
\newcommand{\tmu}{\tilde{\mu}}
\newcommand{\tpsi}{\tilde{\psi}}
\newcommand{\cN}{\mathcal{N}}

\let \t=\tau

\let \vphi=\varphi
\newcommand{\ttmu}{\tilde{\mu}}

\begin{document}
\allowdisplaybreaks

\title{Bose-Einstein condensates in quasi-periodic lattices: bosonic Josephson junction, self-trapping, and multi-mode dynamics}
\author{H. C. Prates} 
\affiliation{Centro de Física Teórica e Computacional and Departamento de Física, Faculdade de Ciências, Universidade de Lisboa, Campo Grande, 1749-016 Lisboa, Portugal\looseness=-1}
\author{D. A. Zezyulin}
\affiliation{School of Physics and Engineering, ITMO University, St. Petersburg 197101, Russia}
\author{V. V. Konotop}
% \email[Corresponding author. E-mail:~]{vvkonotop@fc.ul.pt}
\affiliation{Centro de Física Teórica e Computacional and Departamento de Física, Faculdade de Ciências, Universidade de Lisboa, Campo Grande, 1749-016 Lisboa, Portugal\looseness=-1}
 % \date{November 2021}
\begin{abstract}
   Bose-Einstein condensates loaded in one-dimensional bichromatic optical lattices with constituent sublattices having incommensurate periods is considered. Using the  rational approximations for the incommensurate periods, we show that 
    below the mobility edge the localized states are distributed nearly homogeneously in the space and explore the versatility of such potentials. We show that superposition of symmetric and anti-symmetric localized can be used to simulate various physical dynamical regimes, known to occur in double-well and multi-well traps. As examples, we obtain an alternative realization of a bosonic Josephson junction, whose coherent oscillations display beatings or switching in the weakly nonlinear regime, describe selftrapping and four-mode dynamics, mimicking coherent oscillations and self-trapping in four-well potentials.
    These phenomena can be observed for different pairs of modes, which are localized due to the interference rather than due to a confining trap.   The results obtained using few-mode approximations are compared with the direct numerical simulations of the one-dimensional Gross-Pitaevskii equation.  The localized states and the related dynamics are found to persist for long times even in the repulsive condensates. We also described bifurcations of the families of nonlinear modes, the symmetry breaking and stable minigap solitons.
\end{abstract}
\maketitle

\section{Introduction}

The interaction and evolution of two weakly coupled spatially localized Bose-Einstein condensates (BECs) are one of the basic and well-studied problems in the physics of cold atoms. BECs confined in a double-well trap, also known as a bosonic Josephson junction (BJJ)~\cite{Java,Leggett,GOber}, is the paradigmatic implementation of such a system. It has been observed experimentally~\cite{Albiez2005}, and thoroughly studied theoretically within the quantum and mean-field two-mode models (see e.g.~\cite{Cirac,Steel,SpeSip,Smerzi97,JavIvan,Raghavan1999,Pit2001,ADS,MaPeWilRe,ShTri,Ground2009}). The validity of the two-mode approximation was also discussed~\cite{Anan2006}. Among various fundamental nonlinear phenomena that can be observed in such systems in the meanfield regime we mention coherent oscillations and nonlinear selftrapping~\cite{Smerzi97,Raghavan1999}, spontaneous symmetry breaking~\cite{GarAbd,TheKev}, coupled solitons~\cite{Shchesn2004pra,Shchesn2004physD}, and others.
A common feature of the mentioned theoretical studies is the use of a two-mode approach. In the meanfield approximation, this approach reduces the Gross-Pitaevskii equation (GPE) to a system of two coupled nonlinear ordinary differential equations for the amplitudes of BECs in each of the potential wells. In the one-dimensional (1D) setting the resulting model is also known as a nonlinear dimer~\cite{Kenkre}. 

In the two-mode approximation, the role of a double-well potential is reduced to a confining mechanism for two (or several) low-energy states in each of the potential wells. This results in typical two-hump distributions of two BEC clouds spatially separated by a potential barrier. Specific characteristics of the potentials are accounted for by the coupling coefficient and by the effective nonlinearities. In the meantime, the described confining mechanism, i.e., a double-well trap,  is not the only possibility of creating desirable atomic densities localized in space, even in the linear regime. Indeed, the Anderson localization~\cite{AL,LifGredPas} in a random potential is an example of an alternative physical mechanism enabling spatial localization. However unlike a double-- or multi--well trap, a random potential does not offer an easily controllable way of centering wavepackets at desirable spatial locations or generating pairs of modes of desirable symmetries.

An easier controllable approach could be the use of bichromatic quasi-periodic potentials~\cite{Diener,Modugno,Biddle,BidSar,Yao2019, Boers2007}, say optical lattices characterized by incommensurate periods of the constituent sublattices.  
Furthermore, considered on a sufficiently large, but finite-size interval (as this always occurs in an experiment), a quasi-periodic potential can be viewed as a limit of two commensurate sublattices where the ratio between the periods is a rational approximation (RA) of the irrational number defining quasi-periodicity~\cite{Diener,Modugno,ZeKo2022}. Thus the localization in some sense (specified below) can be achieved even in large enough commensurate superlattices. 

These properties determine the versatility of quasi-periodic potentials, which is explored in our work.
For instance, by a judicious choice of such a potential (or of its commensurate approximation), one can obtain pairs of two-hump even and odd states, which similarly to the states in a double-well potential can be superimposed to provide two weakly localized BECs (an example of similar states was shown in~\cite{BidSar}). The dynamics of these two-hump states should resemble the dynamics of a BEC in a double-well tap. Since a quasi-periodic potential may offer a rich choice of the pairs of localized modes, it also allows to explore the evolution of four (or more) weakly coupled condensates, leading to more sophisticated multi-mode dynamical scenarios. Furthermore, unlike in a double-well trap, in this new setting amplitude variations of the potential are allowed to be relatively small, i.e., the optical lattices used for its creation can be relatively shallow and hence far from the regime of the tight-binding approximation. Quasi-periodic potentials also enable the existence of families of nonlinear modes which can undergo symmetry-breaking bifurcations.  In this work, we describe the above phenomena  which  result from  the interplay between the spatial and energy distributions acting simultaneously, while  
previous publications explored effects based on the features of the energy spectra of quasi-periodic potentials.

The paper is organized as follows. The model and the approach based on the RAs are described in   Sec.~\ref{sec:model}. Characteristics of the linear modes including those allowing for emulation of a BJJ are described in Sec.~\ref{sec:linear}. The two-mode approach for the weakly linear case and the respective BEC dynamics within the framework of the GPE are described in Sec.~\ref{sec:two-mode}. Analysis of the bifurcations of families of stationary nonlinear modes, of the symmetry breaking, and of minigap solitons sustained by an incommensurate bichromatic lattice, is performed in Sec.~\ref{sec:bifurcations}. In Sec.~\ref{sec:4modes} we derive the four-mode Hamiltonian and discuss the evolution of coupled BECs. A possible protocol for preparing the desired localized linear states in a quasi-periodic potential are described in Sec.~\ref{sec:protocol}. The outcomes are summarised in the Conclusion.

\section{The model and its rational approximations} 
\label{sec:model}

Consider a 1D GPE 
\begin{equation}
 \label{eq:GP_nond}
    i \pdv{\Psi}{t}= H\Psi+ g\abs{\Psi}^2 \Psi, \quad H:=- \frac{1}{2}\frac{\partial^2}{\partial x^2}+V(x),  
\end{equation}
for the dimensionless order parameter $\Psi(x,t)$ which is normalized such that $g=+1$ and $g=-1$ correspond to positive and negative scattering lengths of the inter-atomic interactions. We aim at considering  potentials $V(x)$, different from a double-well potentials, but by analogy with the latter ones, featuring eigenstates of the linear eigenvalue problem 
\begin{align}
    \label{H}
    H\tpsi(x)=\tmu \tpsi(x)
\end{align}
(hereafter tilde is used to denote the linear eigenvalues and eigenmodes)
which are localized and either symmetric or anti-symmetric. Such modes correspond to atomic density distributions with two symmetric absolute maxima sufficiently well separated from each other. The formulated task can be implemented using a symmetric, $V(x)=V(-x)$, quasi-periodic potential. To ensure two-hump profiles for the lowest states we require $V(x)$ to have a global maximum at $x=0$ (which splits the atomic density similarly to the central barrier in a double-well trap). One of the simplest implementations of such a potential is a combination of two optical lattices of amplitudes $v_1>0$ and $v_2>0$ and having incommensurate spatial periods, say, $\pi$ and $\pi/\varphi$ where $\varphi$ is an irrational number: 
\begin{equation}
\label{pot-irrat}
    V(x)=V_\varphi(x):=v_1 \cos{(2x)}+v_2 \cos{(2\varphi x)}.
\end{equation}
Thus $V_\varphi(x)$ is a quasi-periodic (alias almost periodic~\cite{Bohr}) function. While a specific choice of $\varphi$ is not essential for the analysis performed below, to be specific  in what follows  we use the golden ratio $\varphi=(1+\sqrt{5})/2$. 

In previous studies~\cite{Modugno,Biddle, BidSar, Yao2019}, it has been shown numerically, that the Hamiltonian $H$ with the potential of the type (\ref{pot-irrat}) may sustain states  which are well localized for energies lower than the mobility edge (ME) $\tmu_{\rm ME}$: $\tmu<\tmu_{\rm ME}$. The peculiarity of our choice (\ref{pot-irrat}) is that such linear eigenstates are double-hump.
Indeed, spatially localized (square-integrable) eigenstates of $H$ (considered on the whole real axis) are nondegenerate. Thus, taking into account that for (\ref{pot-irrat}) a local maximum of $|\tpsi|^2$ cannot be situated  at the origin, the parity symmetry implies that any state must have an even number of density maxima, which are located symmetrically with respect to the origin (see the examples in Fig.~\ref{fig:two} below). Let $\ell_n$ be a distance that separates such maxima of $n$th eigenmode. Suppose also that the size of a real condensate, we denote it by $L$, is finite but much larger than $\ell_n$ for the majority of states with $\tmu_n<\tmu_{\rm ME}$: $L\gg \ell_n$. Then, letting $L$ be sufficiently large but still  finite, one can approximate any almost-periodic function $V_\varphi(x)$ in the interval $x\in[-L,L]$ by a periodic one with any desirable accuracy (this follows directly from Bohr's definition of an almost periodic function~\cite{Bohr}). In our case, such approximation can be achieved by replacing $\varphi$ by its RA $M/N$ where $M$ and $N$ are co-prime integers (notice that the value of $N$ uniquely defines also the respective integer $M$ and {\it vice versa}). Such approximations introduce the set of $N\pi$-periodic potentials (in~\cite{Diener} termed as periodic approximants to the incommensurate potential $V_\varphi(x)$)
\begin{equation}
\label{eq:V_quasi}
    V_N(x):=v_1 \cos(2x)+v_2 \cos\left(2\frac{M}{N} x\right). 
\end{equation}
Obviously, this approximation fails in the vicinity of points $N\pi$ on the real axis but remains sufficiently accurate at $x\in[-L,L]$ if $L\ll N\pi$. Indeed, one estimates
\begin{align}
    |V_N(x)-V_\varphi(x)|\leq 2v_2L\left|\frac{M}{N}-\varphi\right|.  
\end{align}
Thus $|V_N(x)-V_\varphi(x)|\to 0$ for $N\to\infty$ and $L$ fixed. 

The above estimate justifies that being interested in nonlinear  modes localized on a finite interval $[-L/2,L/2]$, one can consider the nonlinear eigenvalue problem, making the anstaz $\Psi(x,t)=e^{-i\mu t}\psi(x)$,
\begin{align}
\label{HN}
    \mu\psi=H_N\psi +g|\psi|^2\psi, \quad H_N:=- \frac{1}{2}\frac{\partial^2}{\partial x^2}+V_N(x)
\end{align}
on the interval $x\in [-N\pi/2,N\pi/2]$ where $N\pi\gg L$, as an approximation for  the stationary GPE with the quasi-periodic potential (\ref{pot-irrat}):
\begin{align}
\label{irrat}
    \mu\psi=H\psi +g|\psi|^2\psi.
    %, \quad H_\varphi:=- \frac{1}{2}\frac{\partial^2}{\partial x^2}+V(x).
\end{align}
Furthermore, since for large dimension of the condensate the effect of the boundaries, i.e., of the vicinity of the points $x=\pm N\pi/2$, on the modes localized in the interval $x\in[-L,L]$ is negligible, in the limit of large $N$, the boundary conditions for the stationary GPE (\ref{HN}) can be chosen at will, provided they do not affect the symmetry of the system. Below we use the periodic boundary conditions 
\begin{align}
    \label{bound}
    \Psi(-N\pi/2,t)=\Psi(N\pi/2,t)\quad    [\psi(-N\pi/2)=\psi(N\pi/2)].
\end{align}
Respectively we consider the auxiliary linear eigenvalue problem
\begin{align}
    \label{linear}
    H_N\tpsi=\tmu%_N
    \tpsi  , \quad \tpsi(-N\pi/2)=\tpsi(N\pi/2) . 
\end{align}
%(respective eigenvalues and eigenmodes are denoted with tilde).
%

One of the advantages of the periodic boundary conditions is that with each bichromatic almost periodic function, like $V_\varphi(x)$ in our case, one can formally associate an infinite lattice considering the periodic potential $V_N(x)$ on the whole real axis. This allows us to interpret the modes of the problem (\ref{linear}) as, and consequently approximate the modes of (\ref{H}) by, the Bloch functions of the periodic potential $V_N(x)$ at the center of the Brillouin zone (i.e., at the zero Bloch quasi-momentum).

As it is customary, to quantify the localization of the solutions  we use inverse participation ratios (IPRs), however considering them separately for portions of the  condensate located at $x<0$ and at $x>0$.
Respectively, we introduce numbers of atoms
\begin{align}
	\cN_\pm=\int_{I_{\pm}^{(N)}} \abs{\Psi(x,t)}^2 d x
\end{align}
on the intervals $I_\pm^{(N)}$ defined as 
\begin{align}
	I_-^{(N)}=[-N\pi/2,0]\qquad I_+^{(N)}=[0, N\pi/2]
\end{align}	
 (the total number of atoms is given by $\cN=\cN_++\cN_-$),
and then introduce the left (`$-$') and right (`$+$')  IPRs
\begin{equation}
	\label{chi}
	\chi_\pm = \frac{1}{\cN_\pm^2}
	\int_{I_{\pm}^{(N)}} \abs{\Psi(x,t)}^4 d x.
\end{equation} 

Likewise, to characterize the position of the modes we use the coordinate of the centers of mass (c.m.) of the atomic clouds in the intervals $I_\pm^{(N)}$:
\begin{equation}\label{eq:COM}
    X_\pm =\frac{1}{\cN_\pm}\int_{I_\pm^{(N)}} x \abs{\Psi(x,t)}^2 d x.
\end{equation}
Obviously, for symmetric and antisymmetric stationary modes $ X_+=-X_-$ (such that the separation distance introduced above is given by $\ell_j= X_+-X_-$) and $\chi_+=\chi_-=2\chi$, where $\chi$ is the conventional participation ratio defined by (\ref{chi}) with integration over the whole interval  $I^{(N)}=[-N\pi/2,N\pi/2]$.  By analogy with the usual criterion, a state is considered localized if $1/\chi_\pm\ll N\pi/2$.

\section{Linear modes}
\label{sec:linear}

\subsection{Localization and memory effect}

\begin{figure}%[h]
	\includegraphics[width=0.49\linewidth]{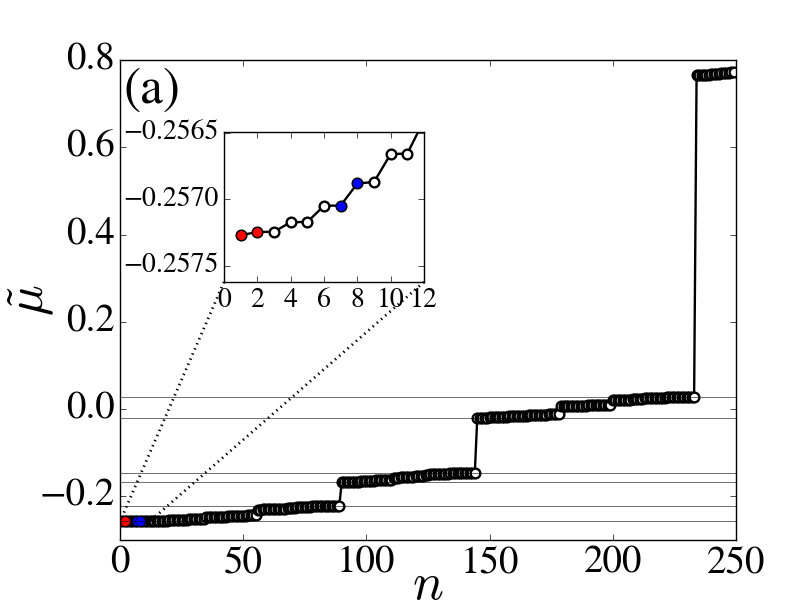}
	\includegraphics[width=0.49\linewidth]{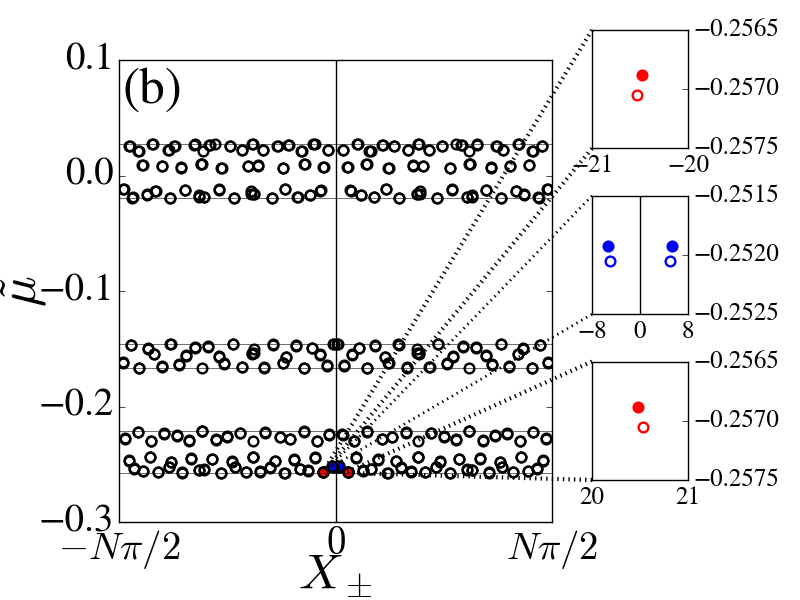}
	\\
	\includegraphics[width=0.49\linewidth]{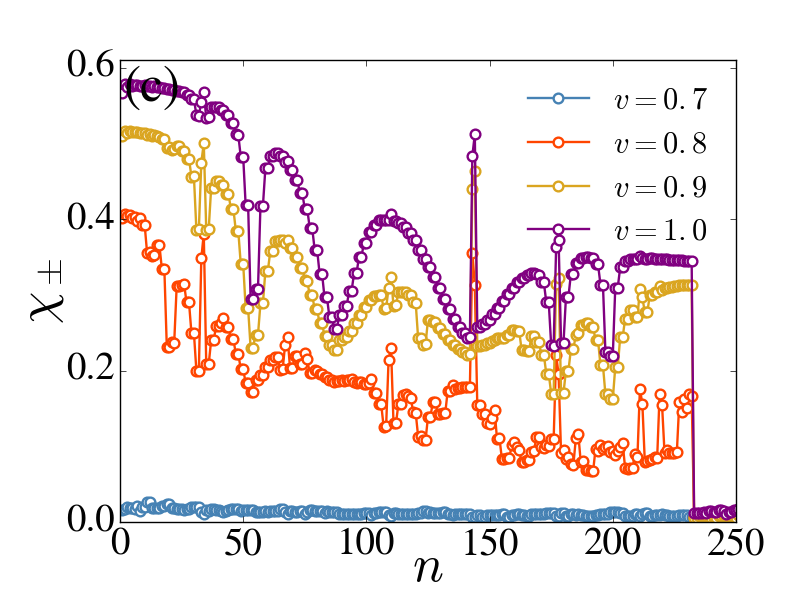}
	\includegraphics[width=0.49\linewidth]{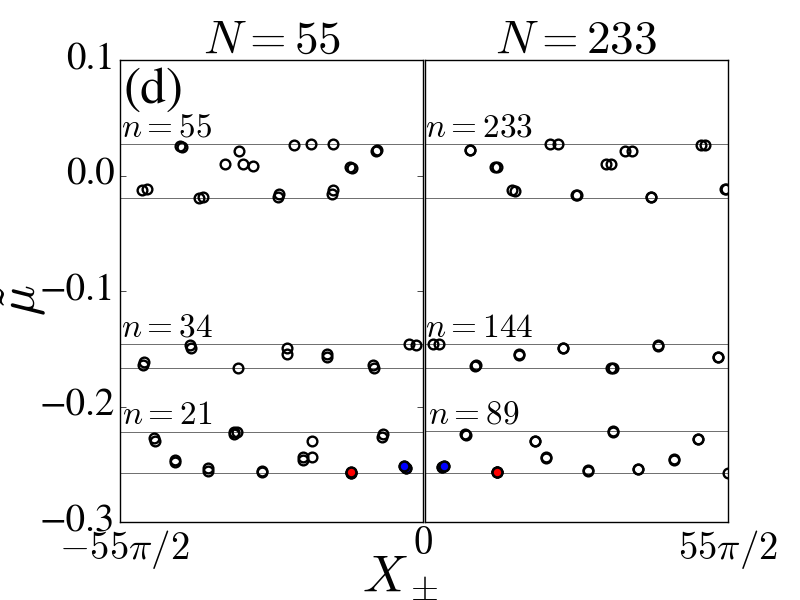}
	\caption{Energies of the 300 lowest linear modes (a), their center of mass coordinates (b) and IPRs (c) for $N=233$ computed  for the RA $N=233$.  The insets in panel (b) show the modes whose explicit shapes are shown in Fig.~\ref{fig:two} and which are used below for observation of the dynamical effects.  In panel (d) we compare the distribution of chemical potentials ans centers of mass for two RA $N=55$ shown for $x\in[-55\pi/2,0]$), and $N=233$ shown for $x\in [0,55\pi/2]$. 
	}
	\label{fig:one}
\end{figure}

We start with the analysis of the eigenmodes of underlying linear problem (\ref{linear}). Since establishing the validity of the approximation of $V_\varphi(x)$ by $V_N(x)$, is relevant for our approach, below we explore two RAs of the golden ratio $\varphi$ with five and six digits of  accuracy, given respectively by $(M_1,N_1)=(89,55)$ and $(M_2,N_2)=(377,233)$. As we will see these approximations are accurate enough to describe the desirable nonlinear phenomena.  The results of the analysis of the linear spectrum are summarized in Fig.~\ref{fig:one}.

In Fig.~\ref{fig:one}(a) we show  the lowest part of the spectrum computed numerically for $N=233$.  We use the lower index $n$ to enumerate the eigenenergies and arrange them   in ascending order: $\tmu_1\leq \tmu_2\leq \ldots$. The shown part of the eigenspectrum  fully covers the localized modes below the sharp ME which is located in the gap between modes with $n=233$ (the highest-energy localized mode) and $n=234$ (the lowest-energy delocalized mode) [cf. with panel (c)]. The spectrum is characterized by groups of eigenvalues separated by minigaps and by the relatively large (alias main) gaps which are located  at the  the modes with numbers $n$ equal to  55,  89, 144, and  233. Location of the largest  gaps can be identified by considering the successive RAs of $\varphi$ since the gaps are open at the center of boundary of the Brillouin zone of the effective periodic potential $V_N(x)$ (this property was discussed in~\cite{Diener}  for the inverse of the golden ratio). Thus, in terms of the spectrum  of the periodic approximant, each RA preserves the memory about the previous (less accurate) ones. In other words, if $N_1<N_2$ then in the approximation $N_2$ one can find an information about the approximation $N_1$.  Indeed, let us start with the spectrum shown in Fig.~\ref{fig:one}(a) which was obtained for $(M_2,N_2)=(377,233)$. The first and the second large  gaps occur, respectively, at the modes with the numbers $N_1=55$ and $M_1=89$. The next large  gap occurs at the mode with the number $N_1+M_1=144$. Notice that $ N_2/(N_1+M_1)=233/144$ is the RA for $\varphi$ between the approximations $N_1$ and $N_2$.  Thus the first three relatively large gaps in the $N_2=233$ approximation occur exactly at the modes whose numbers are obtained from the preceding %$(M_1,N_1)$ 
approximations.

This   {\em `memory effect'} manifests itself not only in the spectrum, but also the real space as discussed below, and it is not restricted to the choice of the golden ratio (say, it was recently observed in a spin-orbit BEC with incommensurate spin-orbit coupling and Zeeman lattice~\cite{ZeKo2022} with the incommensurately relation $\sqrt{2}$). The spatial manifestation of the memory effect is illustrated in Fig.~\ref{fig:one}(d).  Indeed, computing first the c.m. of the localized  modes in the approximation with  $N_1=55$,  we observe that they cover the intervals $I_\pm^{({55})}$ almost homogeneously [since the stationary modes are distributed symmetrically with respect to $x=0$, in  Fig.~\ref{fig:one}(d)  we show only the interval $I_-^{({55})}$]. The modes computed in the $N_2$'s approximation are shown in the interval $I_+^{(55)}$. Recalling that the modes of the problem (\ref{linear}) coincide with the Bloch modes of $V_N(x)$ at  the center of Brillouin zone, one concludes that passing on from the lower $N_1$-approximation to the higher, more accurate  $N_2=233$-approximation   increases the number of modes inside a given interval of the chemical potentials. In the meantime, comparing the left and right parts of Fig.~\ref{fig:two}(d) one can see that c.m. of the modes as well as a number of the modes on the interval $I^{(55)}$ are nearly the same in both approximations. Thus,  the `new' modes that appear when one passes from $N_1$-approximation to $N_2$-approximation,  are almost all localized inside $I^{({233})}$ but outside in the interval $I^{({55})}$ [this part is not shown in Fig.~\ref{fig:one}(d)]. In other words, almost all modes of the $N_2$-approximation inside   $I^{({55})}\subset  I^{({233})}$ are weakly deformed modes of the $N_1$-approximation. Moreover, the mentioned deformation is negligible for modes located far from the interval boundaries as the examples in Fig.~\ref{fig:two} illustrate.

\subsection{Even and odd modes}
 
  In Fig.~\ref{fig:one}(c) we show the half-space IPRs   $\chi_\pm$ obtained for the RA with $N=233$. There exists  a sharp ME located in the energy gap between the modes with numbers $n=233$  and $n=234$: all modes below the ME, i.e., those with $\tmu\leq \tmu_N$, are well localized, and the modes above the ME, i.e., those with   $\tmu > \tmu_N$,    are delocalized. 
%Base on the relation of the memory effect described above on can conjecture that in general the ME of $N$-th RA of the almost periodic bichromatic potential (\ref{pot-irrat}) is located between the modes $N$ and $N+1$. 
The dynamical effects reported below are observed for localized modes below the ME.

\begin{figure}%[h]
       \includegraphics[width=0.49\columnwidth]{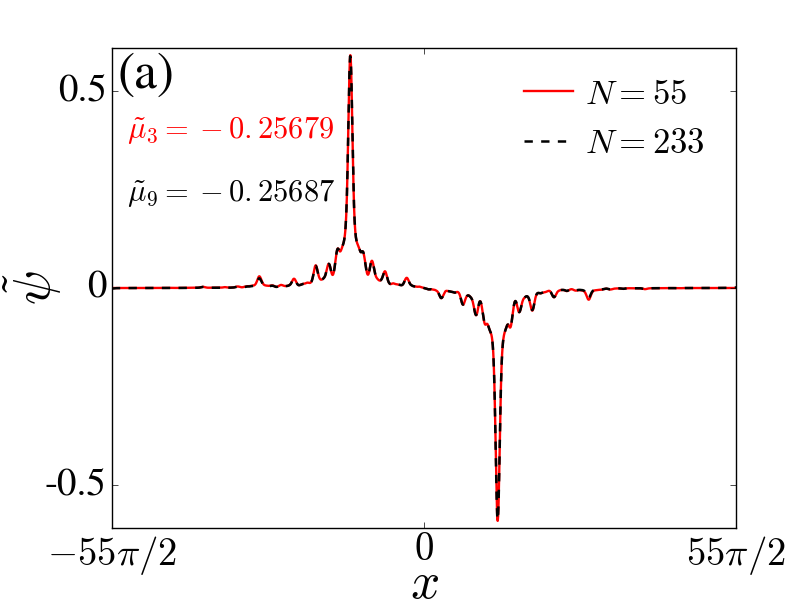}
           \includegraphics[width=0.49\columnwidth]{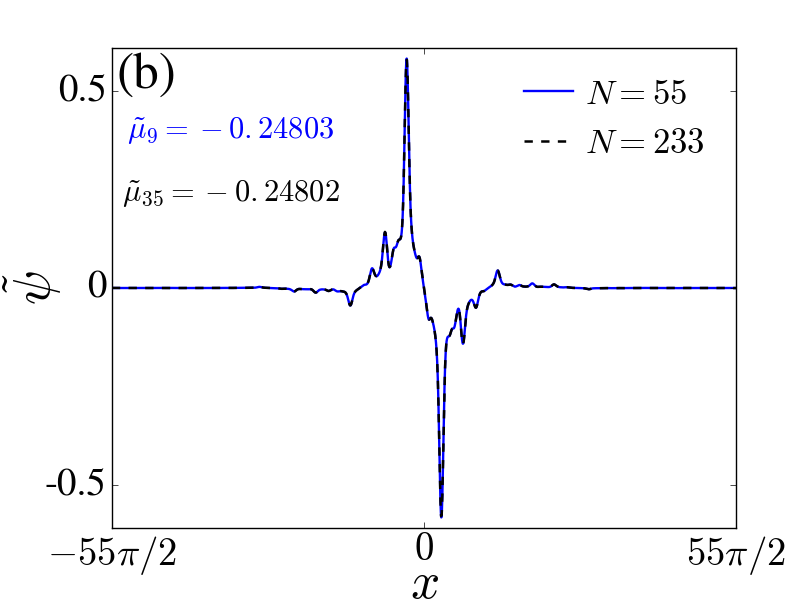}
       \\
          \includegraphics[width=0.49\columnwidth]{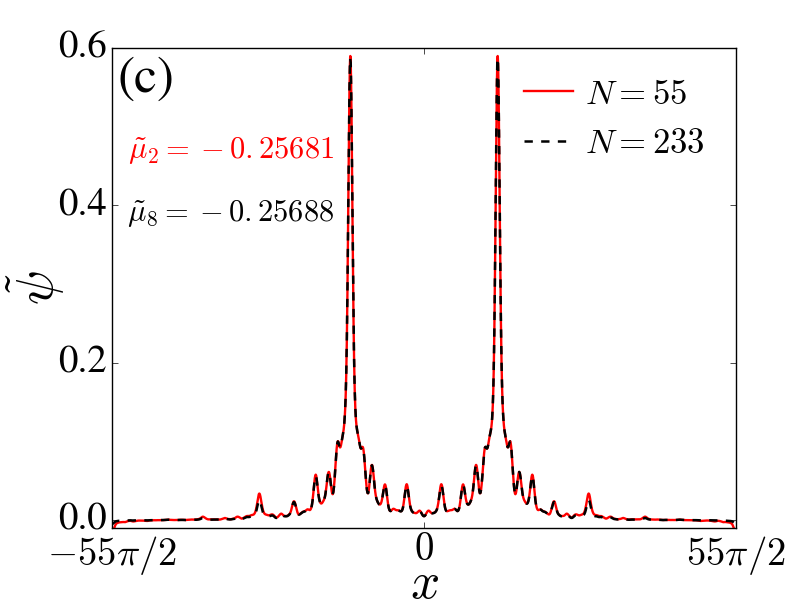}
           \includegraphics[width=0.49\columnwidth]{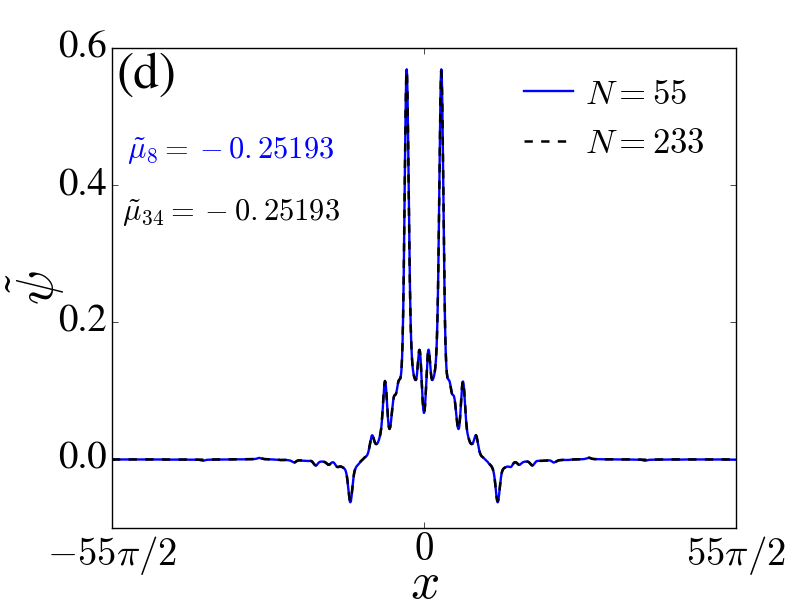}           
 \caption{Two pairs of linear eigenmodes of $H_N$ with $v_1=v_2=0.8$ . In red and blue we plot the modes marked in Fig.~\ref{fig:one} with the same color, for the RA $N=55$, and  dashed lines are used to   represent the corresponding modes for $N=233$.}
     \label{fig:two}
\end{figure}

Due to the symmetry of the potential and location of the absolute maximum at the origin, the modes are either symmetric or anti-symmetric [see Fig.~\ref{fig:one}(d)]. In the case of a quasi-periodic potential considered on the entire real axis, modes vanishing at the infinities are nondegenerate. Therefore, modes of $N$th approximation that are localized sufficiently far from the boundaries are expected to be nondegenerate with the number of nodes increasing with the number of the mode. In other words, symmetric (even) and anti-symmetric (odd) modes should alternate. A relevant numerical finding is that one can identify pairs of successive even and odd modes corresponding to neighbouring (linear) chemical potentials and having maximal absolute values situated at   approximately the same spatial locations (like what happens in a double-well trap). Indeed, Fig.~\ref{fig:two} shows profiles of two pairs of the modes: the lowest two modes $\tpsi_2(x)=\tpsi_2(-x)$  and $\tpsi_3(x)=-\tpsi_3(-x)$ (the left panels) and the modes $\tpsi_8(x)=\tpsi_8(-x)$  and $\tpsi_9(x)=-\tpsi_9(-x)$ (the right panels). The spatial locations of density maxima of these modes with respect to other localized modes in the system are shown by red and blue circles in the insets of Fig.~\ref{fig:one}(b). Symmetric modes shown in Fig.~\ref{fig:two}(c) and (d) are alike an example of a mode symmetric with respect to a potential minimum reported in~\cite{BidSar}, although in our case these modes are even with respect to the potential maximum.

The pairs of the modes like the ones shown in Fig.~\ref{fig:two} allow one to construct the distributions ($j=2,4,...$)
\begin{align}
\label{loc-states}
\varphi_{j}=\frac{\tpsi_j+\tpsi_{j+1}}{\sqrt{2}}, \quad \varphi_{j+1}=\frac{\tpsi_j-\tpsi_{j+1}}{\sqrt{2}} 
\end{align}
which are normalized and mutually orthogonal and are localized mainly in the domains $x<0$ in the point $X_-$ if $j$ is even, and $x>0$ in the point $X_+$ if $j$ is odd. The modes $\varphi_j$ are used below for considering dynamical effects.

Finally, returning to the accuracy of the rational approximants $V_N(x)$, in Fig.~\ref{fig:two} we show all modes for two approximations: $N_1=55$ and $N_2=233$. One can see that already at relatively crude approximation $N_1=55$ the modes are produced numerically with very high accuracy (the distinction between the two approximation is barely visible on the scale of the figure).
 
\section{Bosonic Josephson junction} 
\label{sec:two-mode}

\subsection{Two-mode approximation}

Although the localized modes constructed above share some qualitative features, different pairs can undergo different dynamics stemming from the significant differences in the values of the linear and nonlinear hopping integrals of the modes $\varphi_j$ and $\varphi_{j+1}$. In particular, such integrals rapidly decay with increase of distance $|X_\pm|$ of the modes from the origin. Therefore our first choice is the pair of modes $\varphi_8$ and $\varphi_9$ whose coupling is not negligibly small to ensure coherent oscillations typical for the BJJ [cf. the location of the modes in Fig.~\ref{fig:two}]. Notice that this means that unlike in a double-well trap where the BJJ is usually realized with the two lowest modes, here it will be obtained with higher modes. The approximate analytical treatment of such BJJ, however can replicate the known approach elaborated for a BEC in a double-well trap~\cite{Smerzi97,Raghavan1999}. 
 
 Indeed, consider the two-mode ansatz 
 \begin{align}
\label{two-mode}
 \Psi_a=\left[a_1(t)\varphi_8(x)
      +a_2(t)\varphi_9(x)\right]e^{-i\nu t}
\end{align}
%\begin{align}
%\label{two-mode}
%    \Psi=\left[\sqrt{n_1(t)}e^{i\theta_7(t)}\phi_7(x)
    %\nonumber \\
%    +\sqrt{n_2(t)}e^{i\theta_2(t)}\phi_8(x)\right]e^{-i\nu t}
%\end{align}
where $\nu=(\ttmu_8+\ttmu_9)/2$, in the GPE (\ref{eq:GP_nond}). Since this ansatz implies the approximation $|a_1|^2+|a_2|^2=\cN$ (only two modes are excited) one defines the population $z_a$ imbalance the relative phase $\phi_a$, 
\begin{align}
    \label{za}
    z_a=\frac{|a_1|^2-|a_2|^2}{\cN}, \quad \phi_a=\arg (a_2)-\arg (a_1)
\end{align}
%, where $a_j=\sqrt{n_j}e^{i\theta_j}$ 
thus obtaining 
%(note that  $z\in [-1, 1]$ and $\phi \in [-\pi, \pi)$), 
  the dynamical equations%~\cite{Smerzi97,Raghavan1999}
\begin{equation} 
\label{eq:z_phi}
        \frac{dz_a}{d\t}= -\sqrt{1-z_a^2}\sin\phi_a, \quad 
        \frac{d\phi_a}{d\t}= \frac{z_a \cos \phi_a}{\sqrt{1-z_a^2}}  +  \Lambda_a z_a ,
\end{equation}
where 
%$\Lambda=\frac{U N}{2 \Delta}$. 
%\begin{align}
  $\t=(\tmu_8-\tmu_7)t$ is the re-scaled time,
  \begin{align}
  \label{eq:La}
      \Lambda_a=\frac{g\cN \chi_a}{\tmu_9-\tmu_8},
  \end{align}
 % \Lambda_a=  \frac{g \cN}{{\tmu_8-\tmu_7}} \int_{I} \abs{\varphi_{7,8}}^4 d x  
 and $\chi_a$ is the IPR of any of the $\vphi_8$ and $\vphi_9$ modes.
 Notice that in our case $|a_{1,2}|^2 $ is approximately (but not exactly) equal to $\cN_{+,-}$.   
%, i.e., these quantities are approximately, but not exactly equal to each other, as in the real system other modes not included in the linear two-mode ansatz can become populated.
    %\Delta=\frac{\ttmu_2-\ttmu_1}{2}
    %\\
%\end{align} 
%with $\lambda$ being the wavelength of the laser used to produce the optical lattice and $\ell_\perp$ the transverse diameter of the cigar-shaped BEC.
%Moreover, this can be written in the Hamiltonian form as 
%\begin{equation}
 %   \dot{z}=-\pdv{H}{\phi} \qq{,} \dot{\phi}=\pdv{H}{z},
%\end{equation}
%where 
%\begin{equation}
%    H(z, \phi):=-\frac{g\Lambda z^2}{2}-\sqrt{1-z^2} \cos\phi.
%\end{equation}

\begin{figure}
    \centering
    \includegraphics[width=0.49\linewidth]{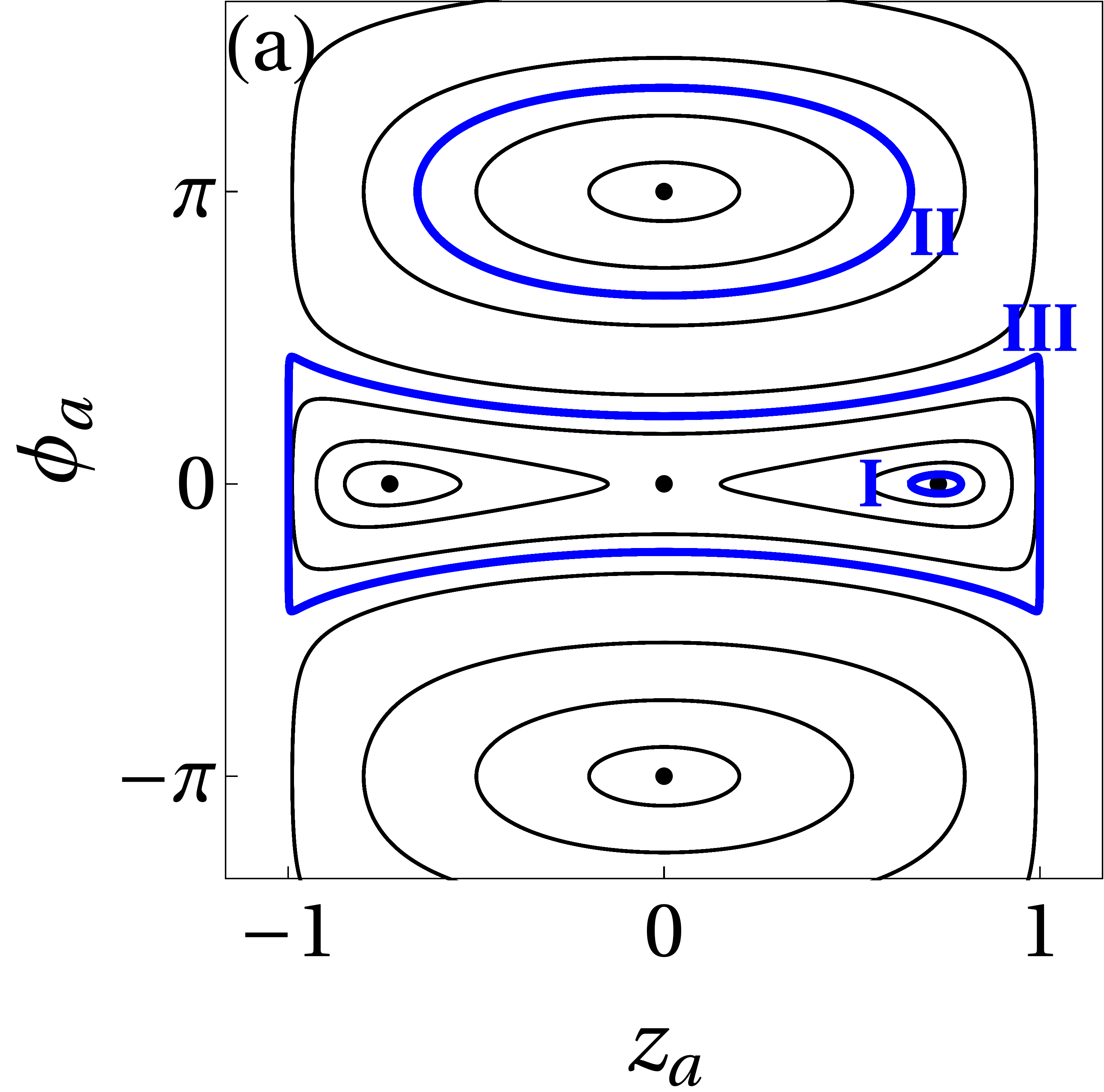}
    \includegraphics[width=0.49\linewidth]{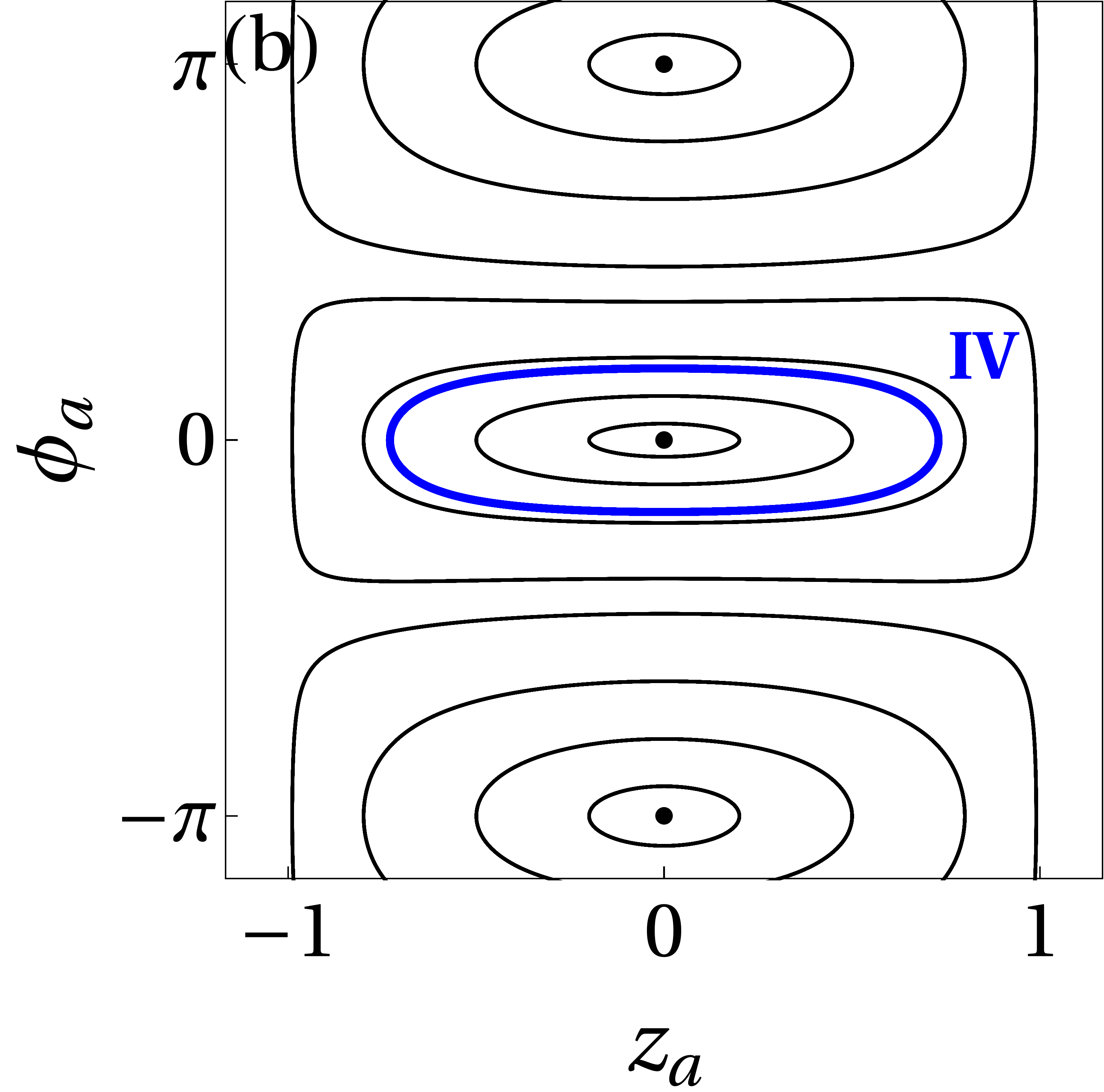}\\
    \includegraphics[width=0.49\linewidth]{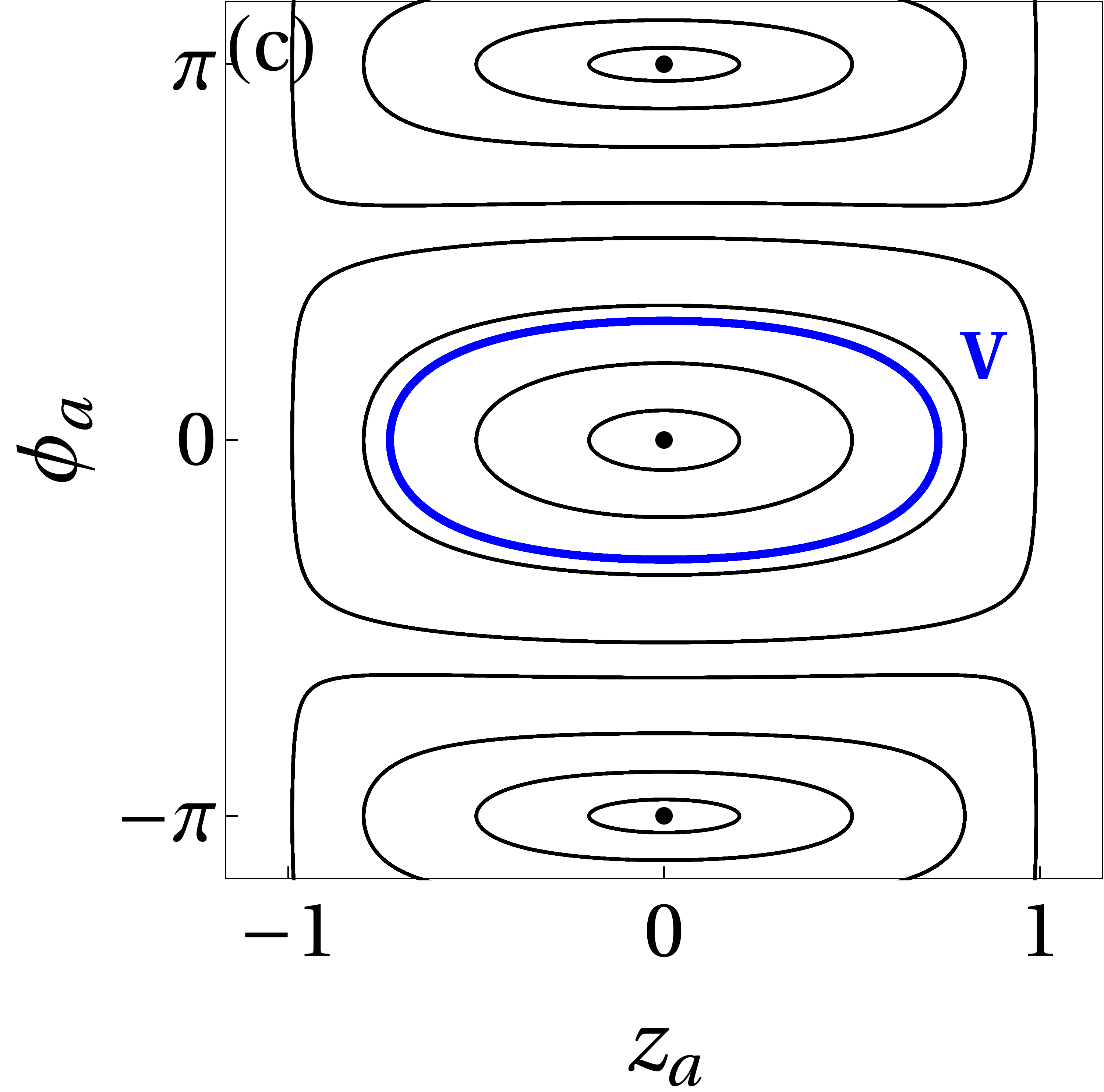}
    \includegraphics[width=0.49\linewidth]{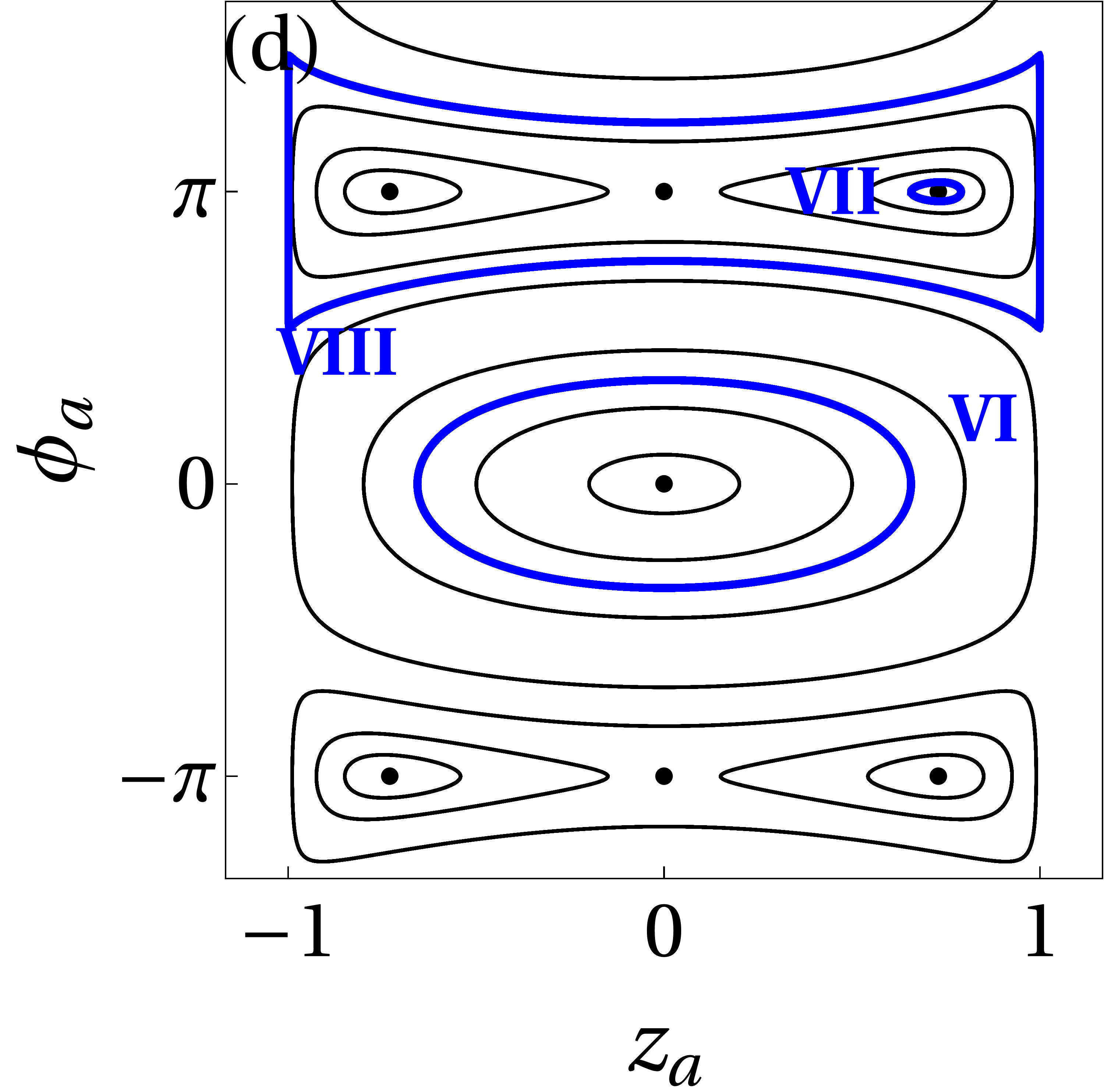}
    \caption{Phase portraits of \eqref{eq:z_phi} for (a) $\Lambda_a <-1$, (b) $-1<\Lambda_a<0$, (c) $0<\Lambda_a<1$, (d) $\Lambda_a>1$. The orbits I-IV and V-VIII correspond to the regimes explored below using direct numerical solutions of the GPE shown in Figs.~\ref{fig:GPE_sol_g_neg} and \ref{fig:GPE_sol_g_pos}.} 
    \label{fig:pportrait}
\end{figure}

The system (\ref{eq:z_phi}) coincides with the one known for a double-well trap~\cite{Smerzi97,Raghavan1999} and its dynamics is well studied. Its phase portraits for different values of the parameters are illustrated in Fig.~\ref{fig:pportrait}. Thus, at the level of two-mode approximation there is no qualitative difference between the dynamical regimes of a BEC in double-well trap and of the chosen couple of modes in a quasi-periodic potential. 
In other words, the considered modes in a quasi-periodic potential emulate a BJJ.
%The difference between the two models, however consists in a possibility of multiple choice of the interacting modes having different spatial locations and thus having different parameters of the two-mode approximation~(\ref{eq:z_phi}). 
However, in contrast to the double-well setup, the quasiperiodic model offers a possibility to consider different pairs of interacting modes that have different locations of absolute maxima and hence   different parameters of the two-mode approximation~(\ref{eq:z_phi}). 

%{\color{blue} Furthermore one needs to consider pairs of modes for which the chemical potential difference $\Delta$ is significant, i.e, $\Delta \gtrsim g \int \abs{\varphi}^4 \dd x$.} 
%Different choices of the modes result in different values of the parameters in the system (\ref{eq:z_phi}). Remarkably, unlike in a double-well trap where the coupling of modes increases with the energy, this is not necessarily so in the case of quasi-periodic potential. For example, coupling of the modes $\varphi_{1}$ and $\varphi_{2}$ constructed using $\tpsi_{1,2}$ (the left column in Fig.~\ref{fig:one}) is much weaker than the that of the modes $\varphi_{8}$ and $\varphi_{9}$ constructed using $\tpsi_{8,9}$ (the right column in Fig.~\ref{fig:one}). Furthermore, considering modes localized sufficiently far from each other .... \ref{Sakaguchi2006}

We illustrate this by comparing the parameters for the modes $\varphi_{8,9}$ which for the sake of brevity we call $a$-modes and for the modes $\varphi_{2,3}$, called below $b$-modes, for which one deduces the  system similar to (\ref{eq:z_phi})  
\begin{equation} 
\label{eq:z_phi_b}
        \frac{dz_b}{d\t}= -\epsilon \sqrt{1-z_b^2}\sin\phi_b, \quad 
        \frac{d\phi_b}{d\t}= \frac{\epsilon z_b \cos \phi_b}{\sqrt{1-z_b^2}}  +  \epsilon \Lambda_b z_b ,
\end{equation}
where $\epsilon=(\tmu_3-\tmu_2)/(\tmu_9-\tmu_8)$, while $z_b$, $\phi_b$ and $\Lambda_b$ are defined for the $b$-modes by analogy with similar definitions for the $a$-modes [see (\ref{za}) and (\ref{eq:La})]. 

As a numerical example, we choose
%The estimates below are made for these parameters correspond to 
$v_1=v_2=0.8$. The respective numerical values of   the chemical potentials are: $\tmu_2=-0.25681$, $\tmu_3=-0.25679$,     $\tmu_8=-0.2519$,  and $\tmu_9=-0.2480$. Thus weakening of the hoping integral due to larger distances between the pumps of the modes, $\ell_{2,3}\gg \ell_{8,9}$ is determined by the factor $\epsilon\approx 0.005 $ greatly slowing down the oscillations of the $b-$modes. As a matter of fact this means that $\varphi_{2,3}$  can be viewed as metastable localized linear eigenstates with anomalously long time of decay.
 
     %\omega=0.00341 \\
    %\Delta_a=1.953 \times 10^{-3} \qq{,} \Delta_b=9.917\times10^{-6}
 
% The evolution of the BECs described by the original GPE (\ref{eq:GP_nond}) is also expected to be different in the case of a quasi-periodic potential, because the localization in a relatively shallow lattice considered here is due to the interference effect rather than due to potential wells confining the condensate. 

\subsection{Negative scattering length}
 \label{sec:neg-scat}
 
Evolution of a BEC described by the original GPE (\ref{eq:GP_nond}) in the case of a quasi-periodic potential is might be different from that of the double-well potential, because the localization in our case  
%a relatively shallow lattice considered here 
is due to the interference effect rather than due to potential wells confining the condensate.
In this and in the next subsection we report the results of the numerical simulations of the the GPE (\ref{eq:GP_nond}) with $V(x)$  replaced by its RA  $V_{55}(x)$ with $N=55$, for the dynamical regimes predicted by the two-mode approximations [see Fig.~\ref{fig:pportrait}].
We use the initial condition as follows
\begin{align}
\label{eq:IC_numerical}
    \Psi(x,0)=\sqrt{\frac{\cN}{2}}\left[\sqrt{1+z_0}\varphi_8(x)+\sqrt{1-z_0}e^{i\phi_0}\varphi_9(x)\right], 
\end{align}
where $z_0=z_a(0)$ and $\phi_0=\phi_a(0)$ are real constants.
 
Starting with the case of a negative scattering length ($g=-1$), we use the parameters of a cigar-shaped (transverse size of a trap $a_\perp=5~ \mu$m) BEC of $^{7}$Li atoms (the scattering length $a_s = -27.6 a_0$, where $a_0$ is the Bohr radius), loaded in an optical lattice of the equal amplitudes  $v_1$ and $v_2$ as mentioned above, which are created by two counter-propagating laser beams with wavelengths $\lambda_1=1~\mu$m and $\lambda_2=(N/M)\lambda_1$. The relation between the norm $\cN$ and the real number of atoms $N_{at}$, controlling the strength of the nonlinearity, is given by $\cN=(2 \lambda_1 a_s)/(\pi a_\perp^2)N_{at}$. The parameters of the dynamical systems (\ref{eq:z_phi}) and (\ref{eq:z_phi_b}) determining the strength of the nonlinearity are $\Lambda_a\approx - 93.18 \cN$ and $\epsilon\Lambda_b\approx-101.88\cN$ (the critical value separating the dynamical regimes for $a-$modes $\Lambda_{a}=-1$, corresponds to 288 $^{7}$Li atoms).

\begin{figure}[h!]
    \centering
    \includegraphics[width=\linewidth]{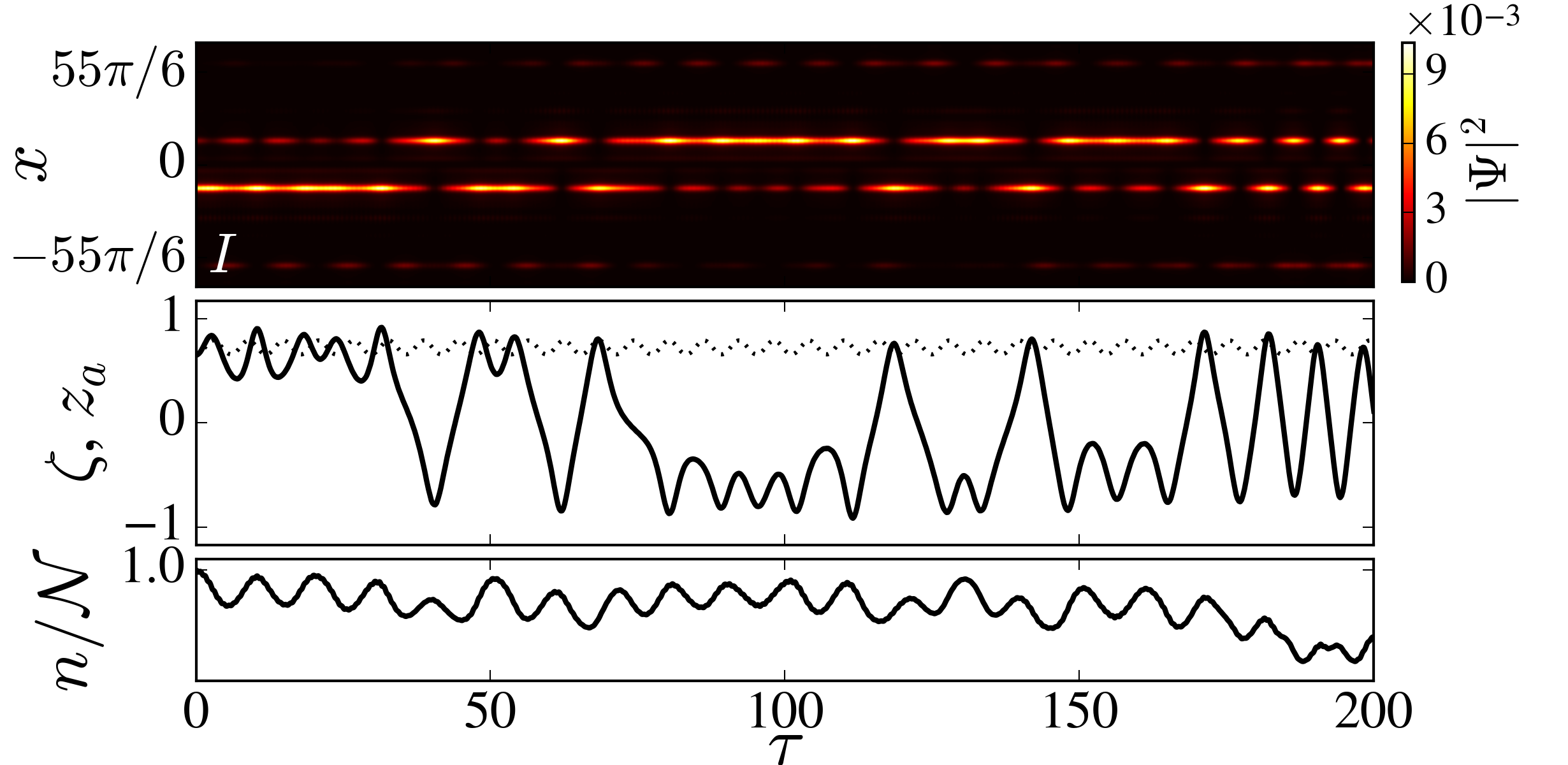}
    \includegraphics[width=\linewidth]{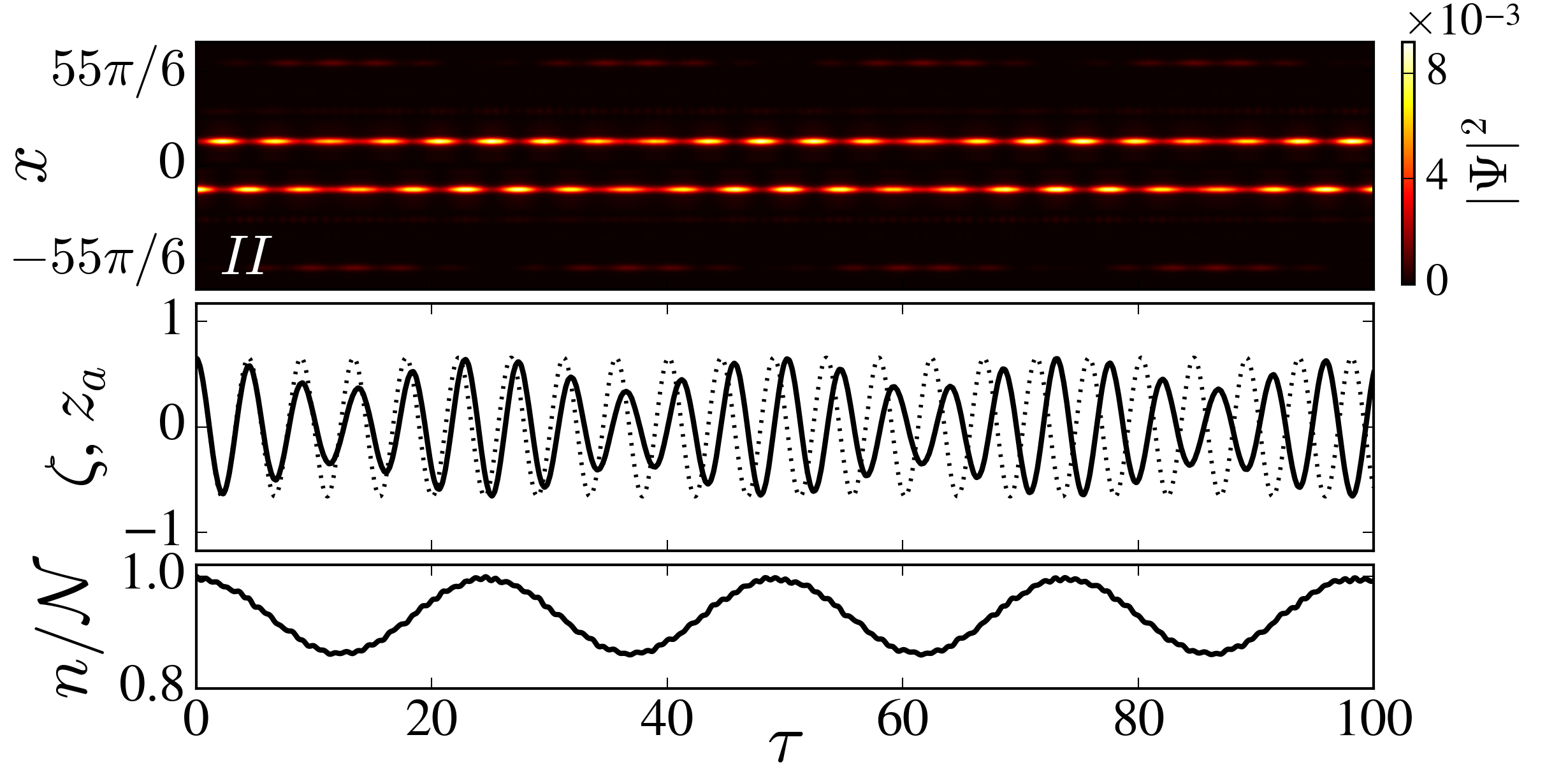}
    \includegraphics[width=\linewidth]{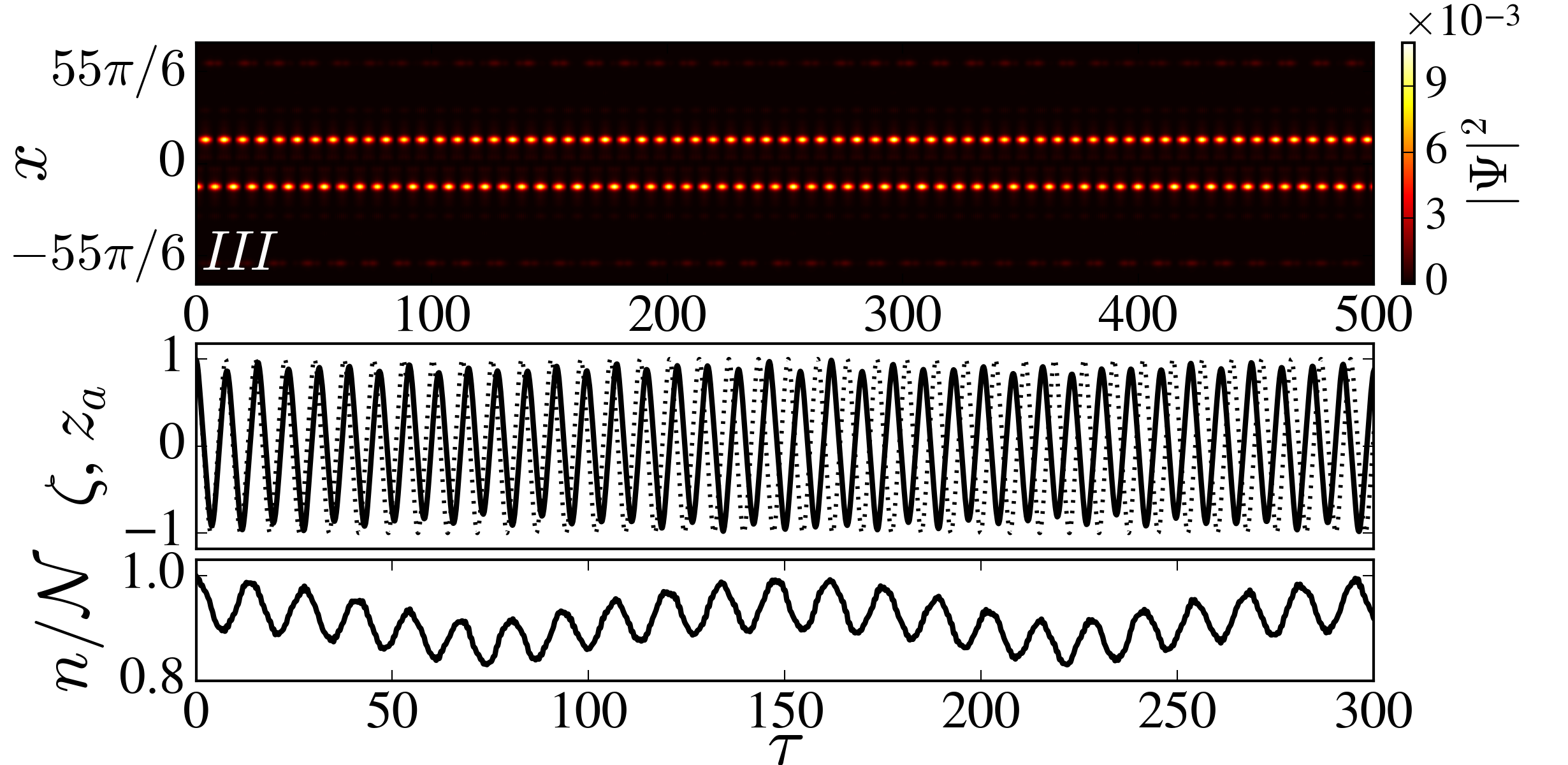}
    \includegraphics[width=\linewidth]{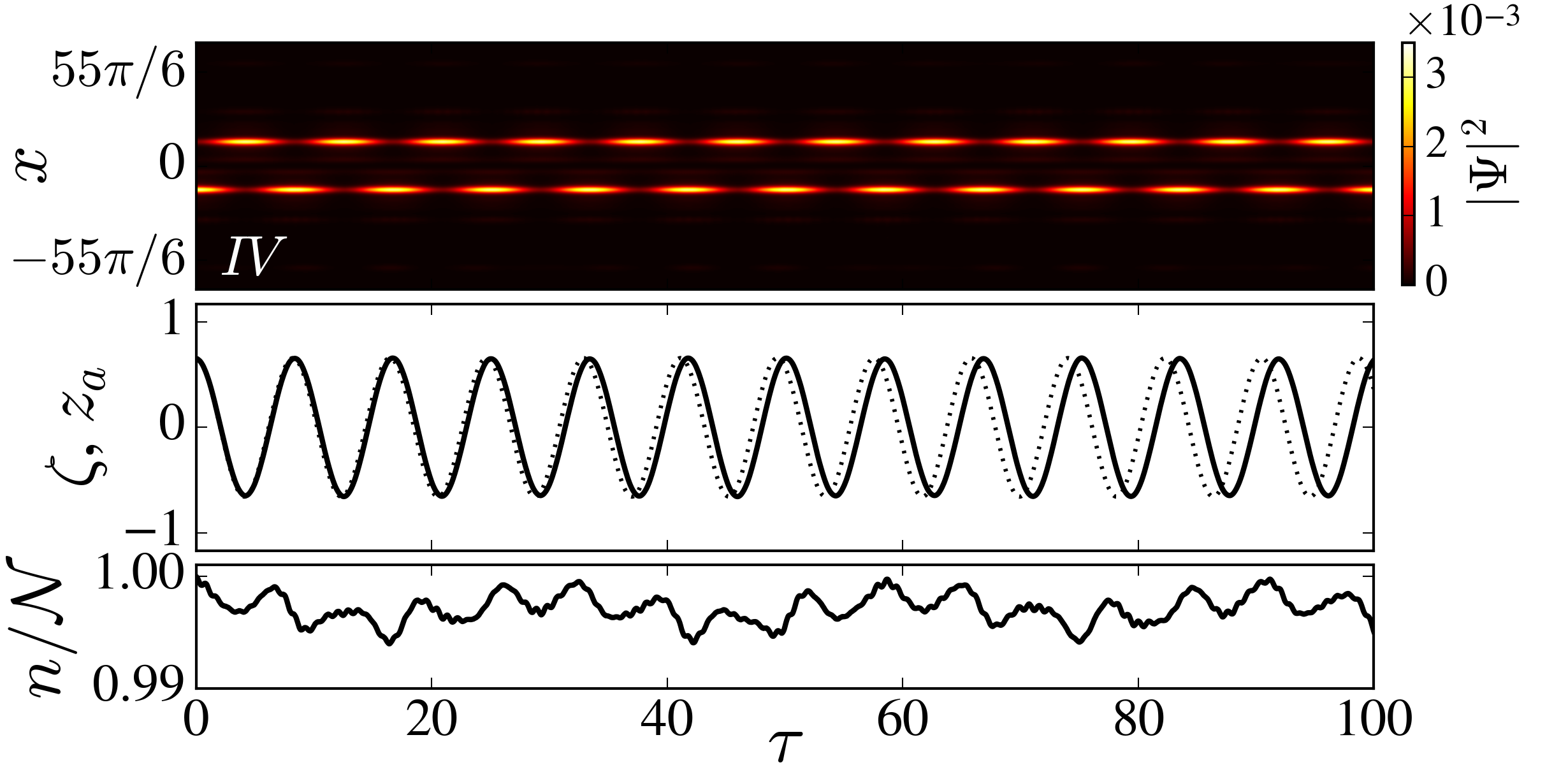}
    \caption{Numerical solutions of GPE \eqref{eq:GP_nond} corresponding to the orbits I-IV in Fig.~\ref{fig:pportrait} with $\Lambda_a=-1.46$ ($\cN=1.57\times 10^{-2}$)
    (I-III) and $\Lambda_a=-0.53$ ($\cN=5.71\times 10^{-3}$) (IV).
    %Each set of panels is indicated by the number of the trajectory. 
    The initial conditions $(z_a, \phi_a)$ are $(0.656, 0)$ for I and IV,  $(0.656, \pi)$ for II, and $(1, 0)$ for III. The upper panels show the density distributions, in the middle panels we show the population imbalance obtained from the direct numerical solution as $\zeta=(\cN_--\cN_+)/\cN$ (solid line) and $z_a$ obtained from the two-mode model (\ref{eq:z_phi}) (dotted line). The lower panels in each set show the fraction of the total number of atoms $n/\cN$ 
    %where $n=|a_1(t)|^2+|a_2(t)|^2$ 
    obtained from the direct simulations using (\ref{eq:n}).
    In the upper panels II-IV showing the full evolution the described coherent oscillations persisted for the whole temporal domain used in simulations: $\tau\lesssim 1200$. }
    \label{fig:GPE_sol_g_neg}
\end{figure}

In Fig.~\ref{fig:GPE_sol_g_neg}I we show the evolution of the initial distribution close to the self-trapping fixed point [trajectory I in Fig.~\ref{fig:pportrait}(a)]. While at the initial stages $\tau$ we observe good quantitative agreement with the two-mode model (cf. solid and dashed lines in the second panel of Fig.~\ref{fig:GPE_sol_g_neg}I), at $\tau\gtrsim 5$ for quantitative and at  $\tau \approx 40$ even qualitative discrepancy of the direct numerical simulations with the two-mode model is observed. Namely,     switching between two fixed points corresponding to self-trapping occurs in the real space: atoms are transferred from the mode $\varphi_8$ (centered at $x<0$) to $\varphi_9$ (centered at $x>0$). The switching occurs several times until the atoms become concentrated in the mode $\varphi_9$. At even longer times $\tau\gtrsim 170$ oscillations corresponding to the trajectory III in Fig.~\ref{fig:pportrait}(a) are established. This failure of the prediction of the full dynamics by the two-mode model can be explained by the excitation of other localized modes upon the evolution governed by the GPE (\ref{eq:GP_nond}) and will be discussed below in Sec.~\ref{sec:switch}. The loss of the atoms by the initially excited states is shown in the lower panel where  the relative number of atoms  
\begin{align}
\label{eq:n}
    n(\tau)=\int\left[\varphi_8^*(x)+\varphi_9^*(x)\right]\Psi(x,\tau)dx
\end{align}
is computed for the $a-$modes (the two mode model corresponds to $n/\cN=1$). The newly excited modes have small amplitudes and are not well visible in the upper panel.
% \begin{figure}[h]
%               \includegraphics[width=0.49\columnwidth]{Figures/phase_I_orb.png}
% %         \caption*{$g=-1.57\times10^{-2}$}
%           \\
%           \includegraphics[width=0.9\columnwidth]{Figures/2w_num/g=-1.57e-02_1_0.png}
%           \includegraphics[width=0.9\columnwidth]{Figures/2w_num/g=-1.57e-02_0.85_pi.png}
%           \includegraphics[width=0.9\columnwidth]{Figures/2w_num/g=-5.71e-03_1_0.png}
%           \includegraphics[width=0.9\columnwidth]{Figures/2w_num/g=-1.00e-01.png}
       
%      \caption{Phase portraits of \eqref{eq:z_phi} and some corresponding solutions of the GPE \eqref{eq:GP_nond}, for the case of negative scattering length. Panels (a) and (e) show the phase portraits for the regimes $g<-\frac{1}{\Lambda}$ and $-\frac{1}{\Lambda}<g<0$, respectively. In (b), (d) and (f) we use as initial conditions $z_0=1$ and $\phi_0=0$, while for (c) we use $z_0=0.85$ and $\phi_0=\pi$. Panel (d) is obtained considering $g\ll -\frac{1}{\Lambda}$.}
     
%     %  Some orbits of \eqref{eq:z_phi} and the corresponding numerical solutions of \eqref{eq:GP_nond}. In panel a) and }
%      \label{fig:three}
% \end{figure}

Stable BJJ oscillations in the full model are also possible and they correspond to the trajectories II -- IV in Fig.~\ref{fig:pportrait}(a), (b) and shown in Figs.~\ref{fig:GPE_sol_g_neg}II -- IV. It turns out, however,  that in the full model, described by GPE, such oscillations are accompanied by beatings of the population imbalance, particularly strong in Figs.~\ref{fig:GPE_sol_g_neg}II, III, and by the frequency shift also visible in Figs.~\ref{fig:GPE_sol_g_neg}IV  [cf. solid and dashed lines in the second panels]. None of this effect is captured by the two-mode model. As above, this can be explained by excitation of the new localized modes, thus resulting in a multimode-dynamics. We also notice that qualitatively, the oscillatory dynamics described by the trajectories II and III within the framework of the two-mode model is different, which is reflected in different beating scenarios (cf. second panels in Figs.~\ref{fig:GPE_sol_g_neg}II and III). The trajectory II is characterized by a simple periodic exchange of the atoms between the original pair of modes and the newly excited ones, while in the case of the trajectory III the population of the original modes undergoes beating itself (cf. the lower panels in Figs.~\ref{fig:GPE_sol_g_neg}II and Figs.~\ref{fig:GPE_sol_g_neg}III). Apart from the described beating and the frequency shifts, the coherent oscillations are sustained by the quasi-periodic potential for a sufficiently long time used in simulations (note the different time scales of the upper and lower panels in Figs.~\ref{fig:GPE_sol_g_neg}III).

In the case of trajectory IV  shown Fig.~\ref{fig:pportrait}(b), the two-mode model accurately describes BJJ (even quantitatively). In Fig.~\ref{fig:GPE_sol_g_neg}IV we observe stable coherent oscillations with a slow increase of the period of oscillations at large times.    

\subsection{Positive scattering length}

\begin{figure}[h!]
    \centering
    \includegraphics[width=\linewidth]{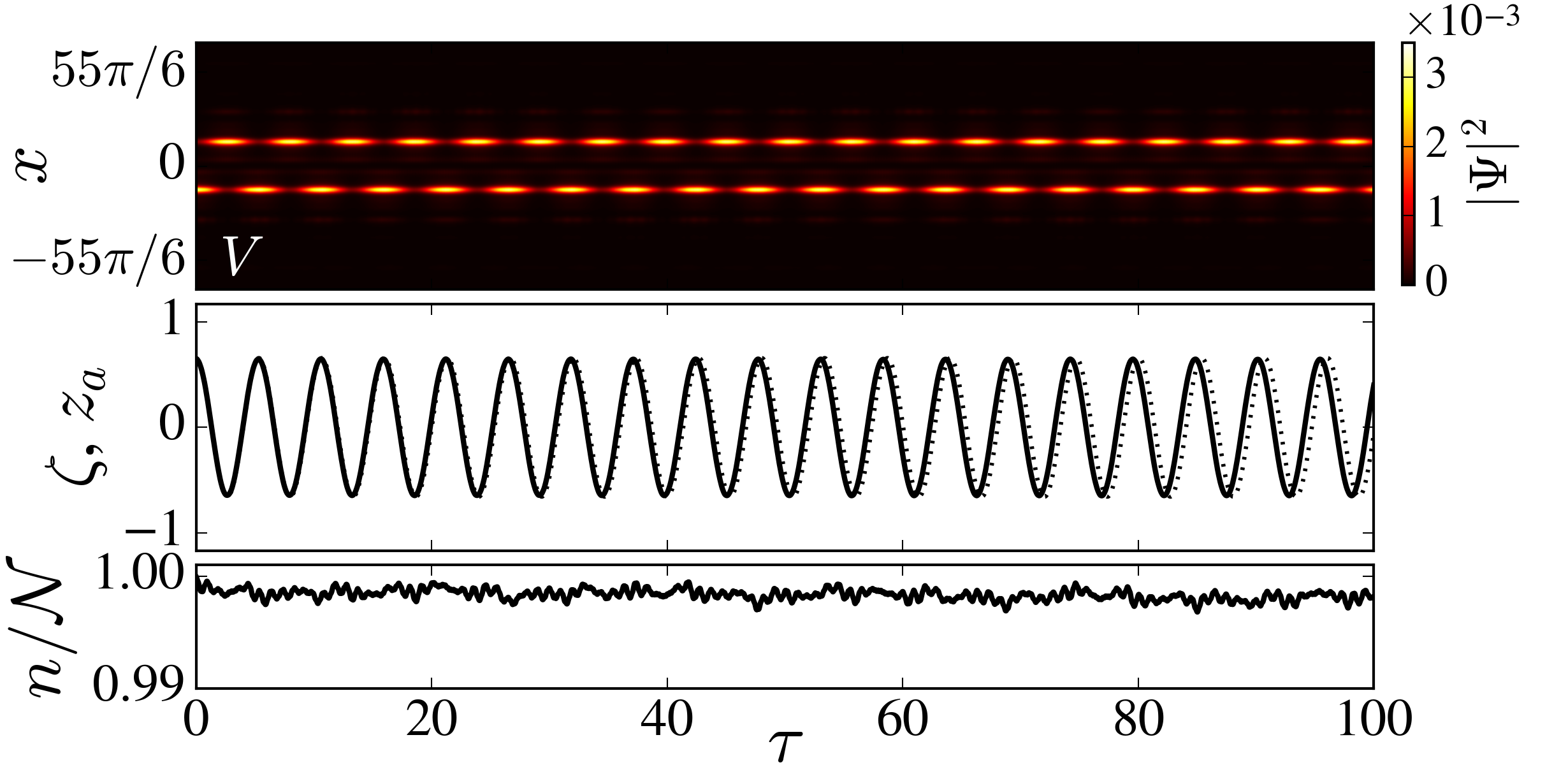}
    \includegraphics[width=\linewidth]{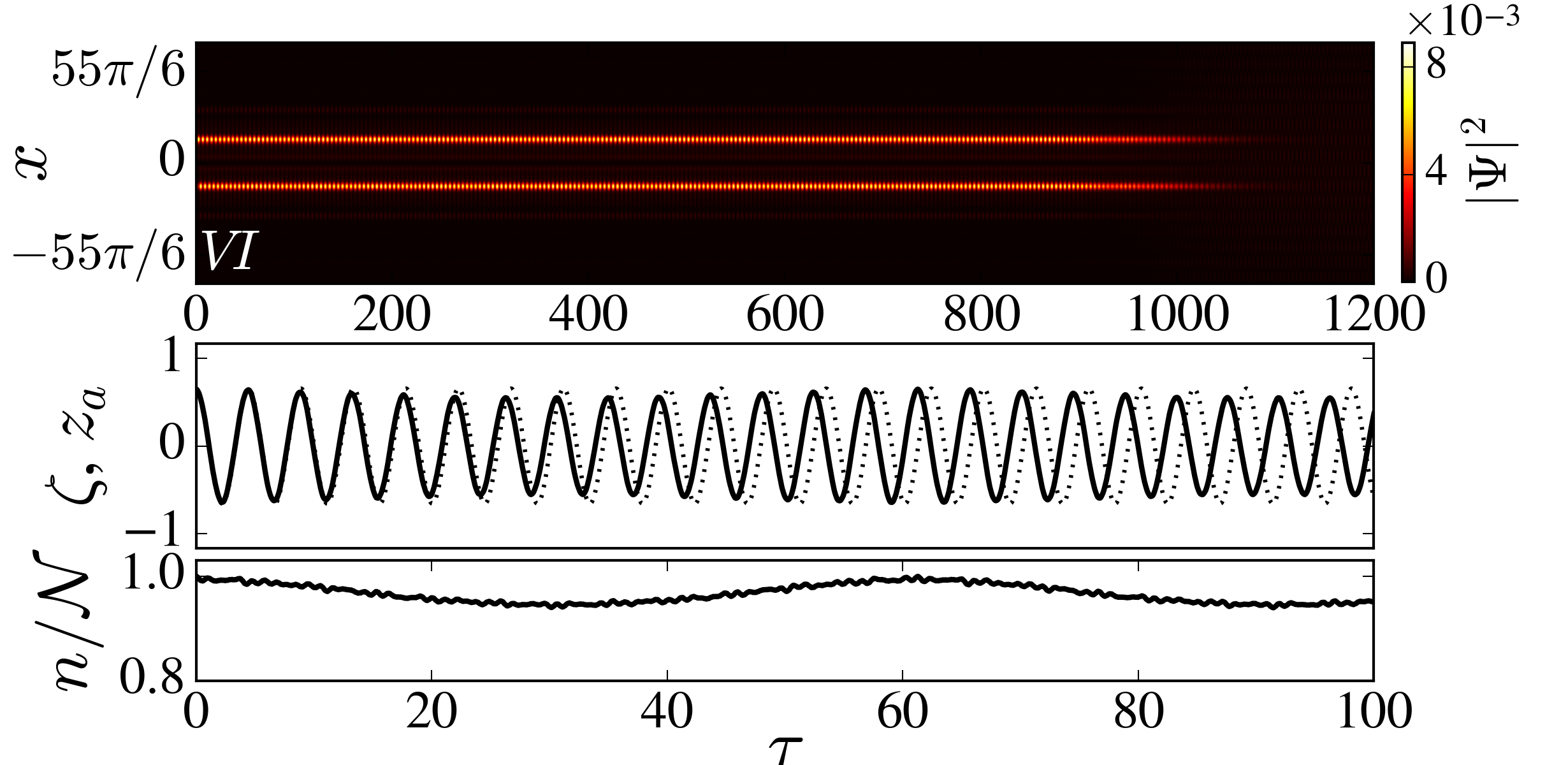}
    \includegraphics[width=\linewidth]{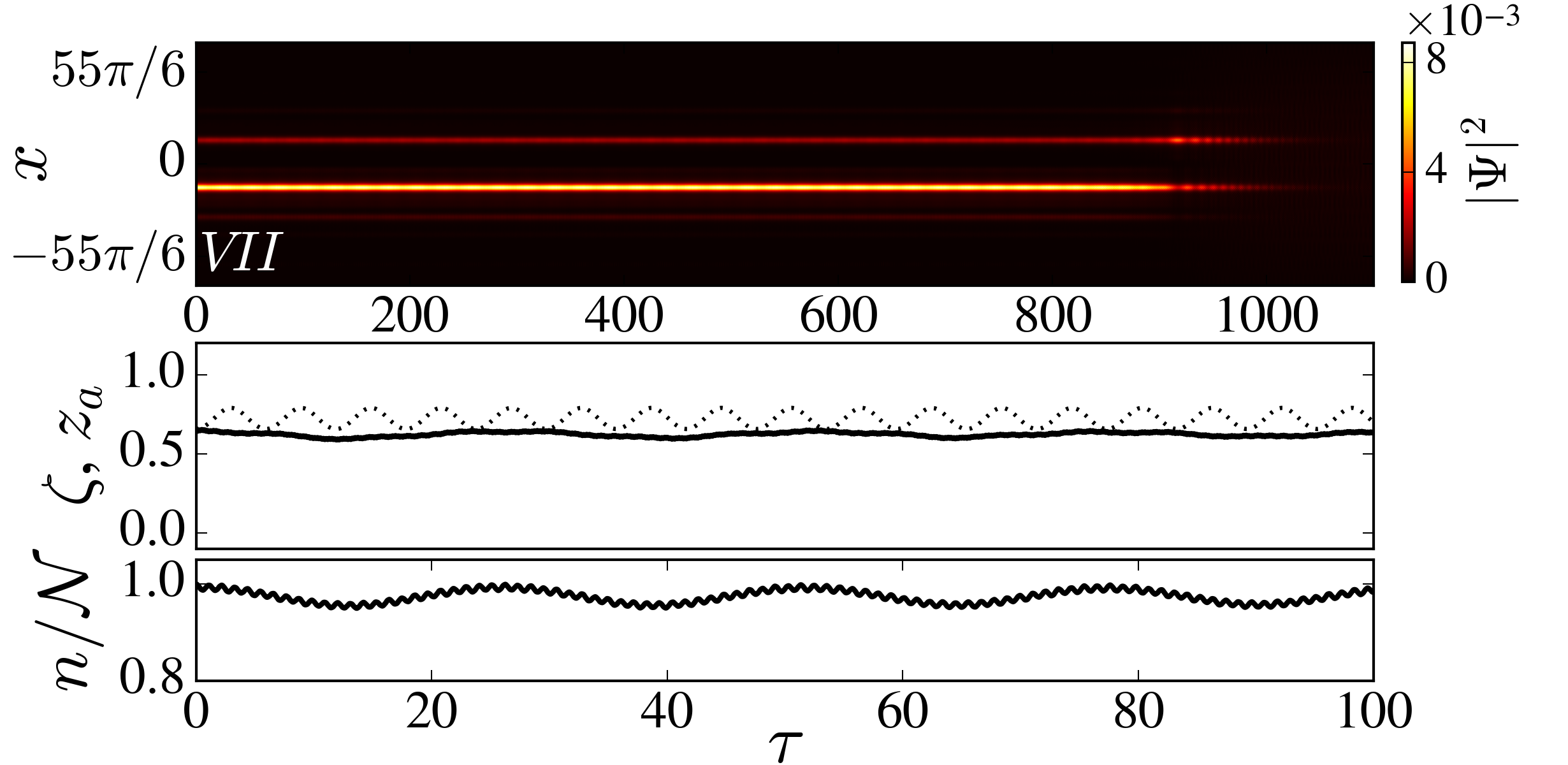}
    \includegraphics[width=\linewidth]{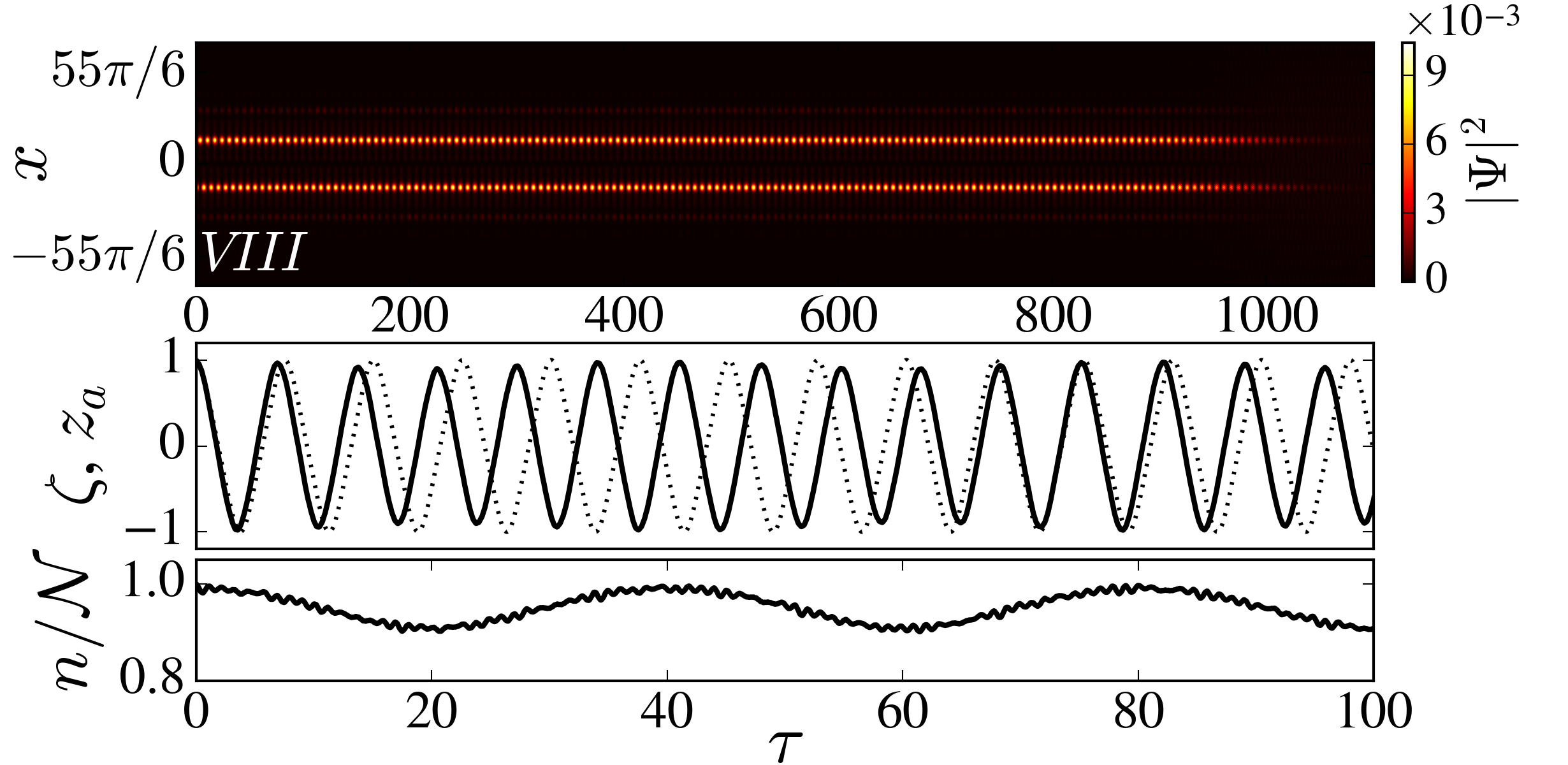}
   \caption{The same as in Fig.~\ref{fig:GPE_sol_g_neg}, but corresponding to the orbits V-VIII in Fig.~\ref{fig:pportrait}, with $\Lambda_a=0.53$ ($\cN=5.71\times 10^{-3}$) (V) and $\Lambda_a=1.46$ ($\cN=1.57\times 10^{-2}$) (VI-VIII). The initial conditions $(z_a, \phi_a)$ are  $(0.656, 0)$  for V and VI, $(0.656, \pi)$ for VII, and $(1, \pi)$ for VIII. Notice the different scales in the upper panels. In the panel V the repulsive inter-atomic interactions are week enough to lead to complete dispersion of the wave-packet on the temporal scales shown in the full evolution in the panels VI-VIII.}
    \label{fig:GPE_sol_g_pos}
\end{figure}
Now we consider BJJ within the framework of the GPE with positive scattering length.  For the sake of definiteness, we refer to  a condensate of $^{87}$Rb atoms, loaded in the same optical lattice  as in the previous subsection. Considering $a_s=95.44 a_0$, we obtain  $\Lambda_a\approx93.18\cN$ and $\epsilon\Lambda_b \approx 101.88 \cN$ (the critical parameter $\Lambda_a=1$ separating the dynamical regimes for the $a-$modes, $\Lambda_a=1$, corresponds to 83 $^{87}$Rb atoms).

Typical results of numerical simulations are summarized in Fig.~\ref{fig:GPE_sol_g_pos}. In Fig.~\ref{fig:GPE_sol_g_pos}V we show coherent BJJ oscillations for times $\tau\lesssim 100$ and observe close  similarities with case of negative scattering length shown in Fig.~\ref{fig:GPE_sol_g_neg}IV. Thus, for the trajectories like V in Fig.~\ref{fig:pportrait}(c), the two-mode model gives an accurate prediction: only a small frequency shift is observed. 
The trajectories VI and VIII in Fig.~\ref{fig:pportrait}(d), in the full dynamics governed by the GPE reveal two phenomena. First, one observes a frequency shift: the GPE solution has a larger frequency compared with the two-mode model. Interestingly, this frequency mismatch between the models is of the opposite sign compared with that observed for negative scattering length [see Fig.~\ref{fig:GPE_sol_g_neg}]. It has an opposite sign also in comparison with the frequency mismatch between the two-mode model and the GPE solution, found previously in~\cite{Anan2006} for the coherent oscillations in a double-well trap. This difference can be related to the fact that here we consider the effect of lower-energy modes 2,3 on the higher-energy modes 8,9.

In addition to the frequency shifts in Fig.~\ref{fig:GPE_sol_g_pos}VI and Fig.~\ref{fig:GPE_sol_g_pos}VIII we observe beating [cf. Fig.~\ref{fig:GPE_sol_g_neg}II and Fig.~\ref{fig:GPE_sol_g_neg}III, respectively]. Like in the case of the negative scattering length considered above, periodic interchange of particles in the two initial states $\varphi_{8,9}$ with newly excited states is expressed in the lower panels where variations of $n/\cN$ of the order of 10\% are illustrated. Meantime, in the repulsive condensate, we did not observe switching of the trajectories in the vicinity of the self-trapping fixed point [Fig.~\ref{fig:pportrait}(d)]. Instead, Fig.~\ref{fig:GPE_sol_g_neg}VII shows a surprisingly (given the repulsive nonlinearity) long-living, the self-trapped state which at $\tau \approx 900$ delocalizes due to the effect of the inter-atomic repulsion. In all the scenarios VI-VIII the evolution ends up with the complete dispersion of the localized atomic states. Meantime we notice that no delocalization was observed for the case shown in  Fig.~\ref{fig:GPE_sol_g_neg}V for the times $\tau\lesssim 1200$.

%\hp{The initial conditions $(z_a, \phi_a)$ for the panels of Figs. \ref{fig:GPE_sol_g_neg} and \ref{fig:GPE_sol_g_pos} are: I-- $(0.656, 0)$, II -- $(0.656, \pi)$, III -- $(1, 0)$, IV -- $(0.656, 0)$, V-- $(0.656, 0)$, VI -- $(0.656, 0)$, VII -- $(0.656, \pi)$, VIII -- $(1, \pi)$.}

% \begin{figure}[h]
    
%           \includegraphics[width=0.49\columnwidth]{Figures/phase_IV_orb.png}
%       %   \caption*{$g=1.57\times10^{-2}$}
%           \includegraphics[width=0.49\columnwidth]{Figures/PIV_z1_phi0.png}
%      %    \caption{$z_0=1$, $\phi_0=0$}
%      \\
%           \includegraphics[width=0.49\columnwidth]{Figures/PIV_z0.85_phi0.png}
%       %   \caption{$z_0=0.85$, $\phi_0=0$}
%          \includegraphics[width=0.49\columnwidth]{Figures/g=1.00e-01.png}
%          \\
%       %   \caption{$z_0=1$, $\phi_0=0$, $g=10^{-1}$}
%             \includegraphics[width=0.49\columnwidth]{Figures/phase_III_orb.png}
%       %   \caption*{$g=5.71\times10^{-3}$}
%           \includegraphics[width=0.49\columnwidth]{Figures/PIII_z1_phi0.png}
%      %    \caption{$z_0=1$, $\phi_0=0$}
    
%      \caption{Phase portraits of \eqref{eq:z_phi} and some corresponding solutions of the GPE \eqref{eq:GP_nond}, for the case of positive scattering length. Panels (a) and (e) show the phase portraits for the regimes $g>\frac{1}{\Lambda}$ and $0<g<-\frac{1}{\Lambda}$, respectively. In (b), (d) and (f) we use as initial conditions $z_0=1$ and $\phi_0=0$, while for (c) we use $z_0=0.85$ and $\phi_0=\pi$. Panel (d) is obtained considering $g\gg \frac{1}{\Lambda}$.}
%      \label{fig:four}
% \end{figure}

\section{Nonlinear stationary solutions}
\label{sec:bifurcations}

\subsection{Families of modes and symmetry breaking}

The two-mode approximation developed above predicts the existence of stationary self-trapping regimes which correspond to fixed points of dynamical systems  (\ref{eq:z_phi}) and (\ref{eq:z_phi_b}). This prediction, however, is based on weakly nonlinear approximations expressed by the ansatz (\ref{two-mode}). In this section, we explore the corresponding stationary modes in the framework of the stationary GPE model.  These regimes can be identified as   eigenstates of the  nonlinear problem (\ref{HN}) with periodic boundary conditions (\ref{bound}).
%$\psi(-N\pi/2) = \psi(N\pi/2)$. 
In the small-amplitude limit $|\psi(x)|\ll 1$ 
%the nonlinear term in (\ref{HN}) becomes negligible, which suggests that  
families of the nonlinear modes bifurcate from the underlying linear problem (\ref{linear}). Standard perturbation theory shows that for $|\mu-\tmu_n|\ll 1$ the shape of a nonlinear mode  $\psi(x)$ bifurcating from a linear mode $\tpsi(x)$ [the latter solves (\ref{linear})] can be approximated as  
\begin{equation}
\label{approx}
    \psi(x) \approx \left(\frac{\mu-\tmu}{g\tilde{\chi}}\right)^{1/2}\tpsi(x),
\end{equation}
where $\tilde{\chi} = \int_{I^{(N)}} \tpsi^4 dx$ is the IPR  corresponding to the underlying linear mode (recall that the linear modes are normalized as $\int_{I^{(N)}} \tpsi^2dx = 1$).  Using this approximation, one can perform the numerical continuation to obtain nonlinear modes of gradually increasing amplitude.

\begin{figure}%[h!]
    \centering
    \includegraphics[width=0.99\linewidth]{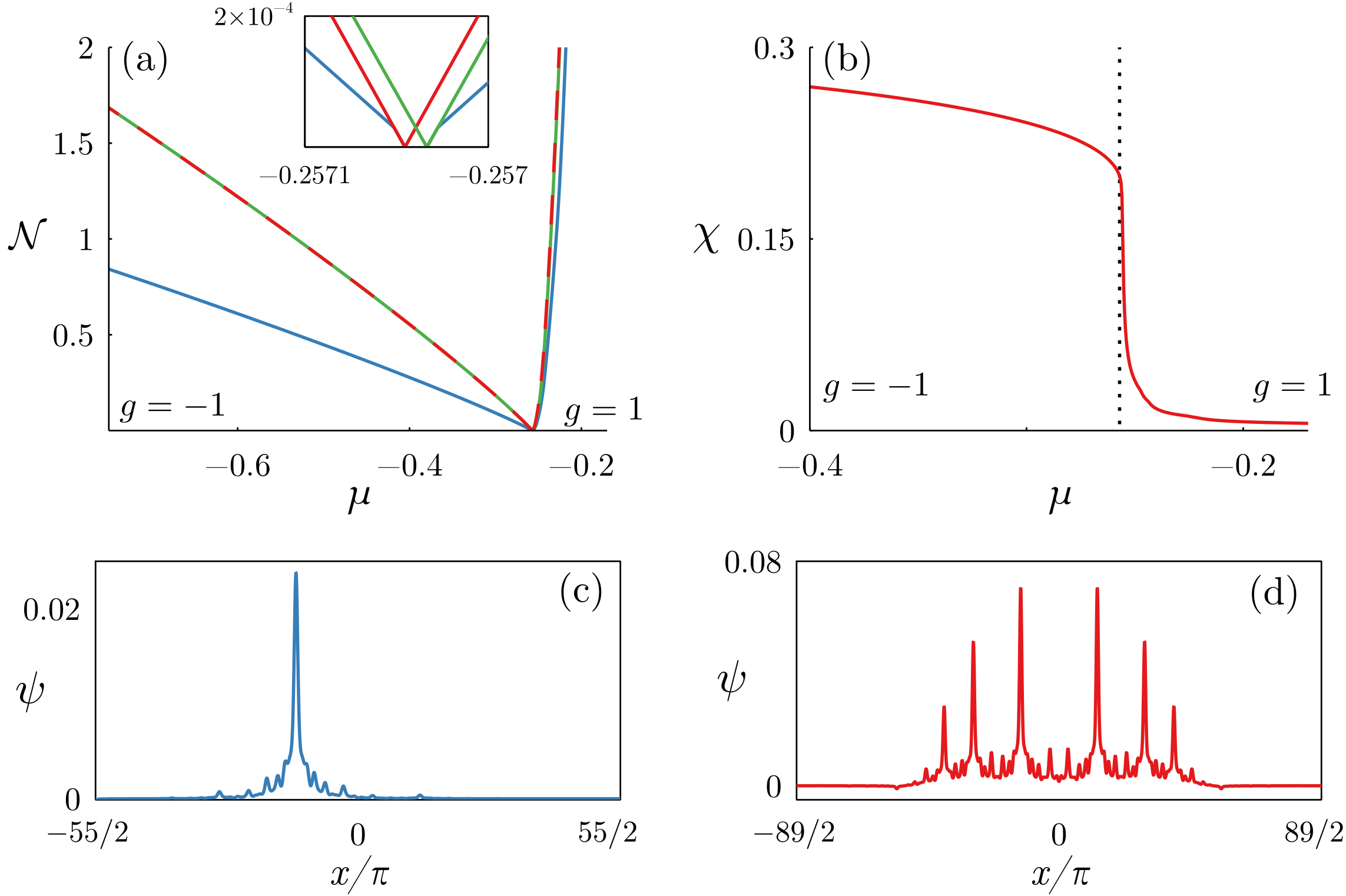}
   \caption{Stationary nonlinear states bifurcating from the linear modes shown in left panels of Fig.~\ref{fig:two}. (a) Dependencies $\cN(\mu)$ computed for even (red color), odd (green color), and asymmetric (blue color) nonlinear modes under attractive ($g=-1$) and repulsive ($g=1$) nonlinearity. The inset  zooms in the region corresponding to the small-amplitude solutions bifurcating from the limit $\cN \to  0$. On the scale of the   main panel, the dependencies $\cN$ for even and odd modes are indistinguishable and therefore presented by dashed lines with alternating colors.  (b) The corresponding dependencies  of IPRs $\chi(\mu)$ for nonlinear even modes  under attractive ($g=-1$) and repulsive ($g=1$) nonlinearity; vertical dashed line corresponds to the $\mu=\tmu$. (c) Profile of asymmetric mode at $\mu=-0.3$. (d) Profile of a symmetric mode under the repulsive nonlinearity at $\mu\approx  -0.2543$. (In c and d only the central part of the total interval $I^{(233)}$ is shown)  }
    \label{fig:nlmodes}
\end{figure}

First, we illustrate nonlinear solutions  bifurcating from  the  pair of linear modes  shown in two left panels of Fig.~\ref{fig:two}, i.e., from $b$-modes in terms of Sec.~\ref{sec:two-mode}. In our simulations we use the  rational approximation with $M=377$ and $N=233$. Figure~\ref{fig:nlmodes}(a) presents the dependencies $\cN(\mu)$ for different families of nonlinear modes. Attractive ($g=-1$) and repulsive ($g=1$) nonlinearities correspond to dependencies with negative and positive slopes, respectively. Regarding the spatial shape of nonlinear modes, approximation (\ref{approx}) implies that for either sign of the nonlinearity in the small-amplitude limit the profile $\psi(x)$  is determined by the underlying linear mode $\tpsi(x)$. In the meantime,   the numerical continuation to the region of stronger amplitudes indicates that the impact of attractive and repulsive nonlinearity is significantly different. In the attractive case ($g=-1$) the shape of nonlinear modes remains close to that of the underlying linear mode. As a result, the IPR of nonlinear modes [defined as $\chi = \cN^{-2} \int_{I^{(N)}} \psi^4 dx$ and plotted  in Fig.~\ref{fig:nlmodes}(b)] does not deviate significantly from the IPR of the underlying linear mode: the IPR only slightly increases with the growth of $\cN$ (i.e., with the decrease of $\mu$ towards the negative infinity) which is a natural consequence  of the attractive self-action.   However,   the repulsive nonlinearity ($g=1$)   excites multiple  new   densely located  peaks, and the increasing number of these new peaks  eventually leads to delocalization of the nonlinear mode whose IPR decays monotonously and eventually becomes very  close to zero, see the plot with $g=1$ in Fig.~\ref{fig:nlmodes}(b) and an example of symmetric nonlinear mode with multiple new peaks plotted in Fig.~\ref{fig:nlmodes}(d).

Apart from even and odd nonlinear states bifurcating from  the linear modes of   corresponding parity, we have also found asymmetric modes  emerging as a result of a symmetry-breaking bifurcation that occurs for  families of  even and odd modes in  the attractive and repulsive cases, respectively. Clearly, each asymmetric state $\psi(x)$ has a partner state given by $\psi(-x)$, i.e., the bifurcations are   pitchfork-type. The symmetry-breaking bifurcations occur for small but nonzero threshold values of the total number of particles $\cN$   and are well visible in the inset in Fig.~\ref{fig:nlmodes}(a).  As follows from the double-mode approximation developed above, the moment of symmetry-breaking bifurcation  corresponds to the condition $|\Lambda_b|=1$, and therefore the smallness of the threshold value $\cN$ is determined by the smallness of the difference $|\tmu_1-\tmu_2|$. The observed symmetry-breaking bifurcations strongly resemble those occurring in the double-well potentials, where the lowest (even) and first excited (odd) nonlinear states are known to undergo  supercritical  symmetry-breaking bifurcations in the attractive and repulsive cases, respectively, see e.g. \cite{TheKev,SBB1,SBB2,SBB3,SBB5}. %\vk{\em Dima, perhaps now we have much lower threshold for the bifurcation as compared with the double well. Is this true? If "yes" then this should be the effect of the very weak linear coupling $\Delta_{a,b}$. Can we discuss this somehow?"}

\begin{figure}%[h!]
    \centering
    \includegraphics[width=0.99\linewidth]{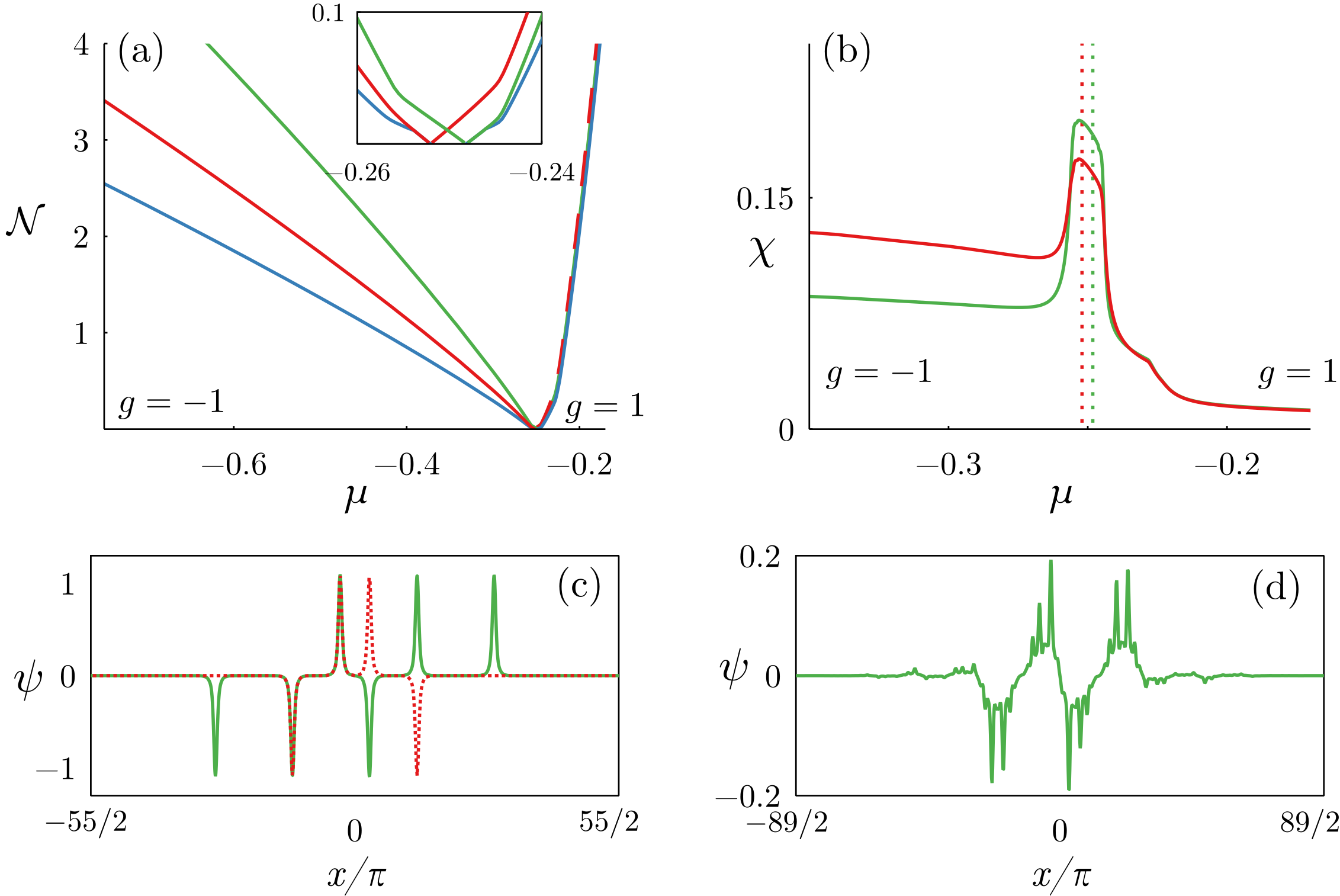}
   \caption{Stationary nonlinear states bifurcating from the linear modes shown in right panels of Fig.~\ref{fig:two}. (a) Dependencies $\cN(\mu)$ computed for even (red color), asymmetric (blue color), and  odd (green color)  nonlinear modes under attractive ($g=-1$) and repulsive ($g=1$) nonlinearity. The inset  zooms in the region corresponding to the small-amplitude solutions bifurcating from the limit $\cN \to 0$.  (b) The corresponding dependencies of IPRs $\chi(\mu)$ for nonlinear even and odd modes  under attractive ($g=-1$) and repulsive ($g=1$) nonlinearity; vertical dashed lines correspond  to the $\mu=\tmu$. (c) Profile of even and odd     modes at $\mu=-1$. (d) Profile of an odd mode under the repulsive nonlinearity at $\mu\approx  -0.2283$. (In c and d only the central part of the total interval $I^{(233)}$ is shown).}
    \label{fig:nlmodes2}
\end{figure}

For nonlinear solutions bifurcating from linear modes shown in the right panels of  Fig.~\ref{fig:two} (i.e., for $a$-modes), the resulting figure becomes slightly different. As shown in Fig.~\ref{fig:nlmodes2}(c), under the attractive nonlinearity the initially two-peaked profiles eventually acquire  several well-separated  peaks, so that the nonlinear profiles  become multi-peaked (with different number of density peaks for even and odd modes). The moment of emergence of new peaks corresponds to a sharp decay of the IPRs plotted  in Fig.~\ref{fig:nlmodes2}(b) for $g=-1$. However,  even after the step decrease, the IPRs of attractive nonlinear modes remain distinctively different from zero. As a result, under the attractive nonlinearity the IPRs feature a counterintuitive, nonmonotonic behavior. The emergence of `new' peaks is related to the fact that as the chemical potential decreases towards the negative infinity, the families of nonlinear modes, bifurcating from $\tmu_7$ and $\tmu_8$, come past the lower chemical potentials (respectively,   the positions of the `new' peaks  are determined by   positions of the density peaks of the eigenfunctions corresponding to the lower chemical potentials).  

The repulsive nonlinearity in Fig.~\ref{fig:nlmodes2} again leads to the emergence of multiple densely located peaks in the profiles of stationary modes.  With the growth of $\cN$ (i.e., with the increase of the chemical potential $\mu$), the newly emerging peaks spread all over the  entire spatial interval $I^{(N)}$ and, as a result, IPRs decrease monotonously and eventually become  very close to zero as shown in Fig.~\ref{fig:nlmodes2}(b) with $g=1$. An example of nonlinear mode which already contains     multiple   newly excited peaks but is  not yet strongly  delocalized is shown in Fig.~\ref{fig:nlmodes2}(d). 

 \subsection{Minigap solitons}
 
Under the continuous change of the chemical potential, the families of localized modes cross several minigaps and minibands. Therefore the respective stationary solutions, considered in the truly aperiodic medium, in our case corresponding to the limit $N\to\infty$,  can be termed to as \emph{minigap} solitons. Such solitons differ from the so-called gap-solitons in quasi-periodic potentials found previously in~\cite{Sakaguchi2006} (and in the aperiodic discrete models~\cite{Johan1,Johan2}), which were obtained for the main gaps, thus resembling the conventional matter gap solitons in periodic media (see e.g.~\cite{BraKon}). Unlike gap bright solitons, bifurcating from the Bloch state at an edge of a linear band towards a gap, the minigap solitons bifurcate from the localized linear modes below the ME and their chemical potentials  are not required to belong  to a spectral gap. Meantime, by analogy to gap solitons, minigap solitons can be found for both positive and negative scattering length [see families in Fig.~\ref{fig:nlmodes2}(a)].

\section{Four-mode dynamics}
\label{sec:4modes}

As we have shown above for BJJ and for nonlinear stationary solutions, when two localized modes are initially excited, the nonlinearity excites other localized states. The effect of such modes on the BJJ is clearly detectable, even though the new excited modes could remain small enough during the evolution. This raises a question about the properties of the {\em multi-mode} dynamics of a condensate in a quasi-periodic lattice. In this section, we consider the evolution of four coupled modes and show that already such an extension of the theory allows one to explain several phenomena observed in Figs.~\ref{fig:GPE_sol_g_neg} and \ref{fig:GPE_sol_g_pos}, but not captured by the two-mode model. We notice that the use of models beyond two-mode approximation, allowing for more accurate accounting the properties of the BJJ in a double-well trap has been reported before~\cite{Anan2006}. There is however several significant distinctions between the modes described below in this Section and those in the double-well trap. The modes considered here are spatially separated (in our case $\ell_2\gg\ell_8$), i.e., the impact of nonlinear interactions is different. We consider the correction of the dynamics of higher energy $a$-modes by accounting for lower energy $b$-modes. Furthermore, below we present the explicit Hamiltonian for the four-mode evolution, allowing, for example, for the exact (withing the four-mode dynamics) analytical computation of the fixed points.

To be specific we explore the case when all four modes $\varphi_{2,3,8,9}$, shown in Fig.~\ref{fig:two} are excited. The respective four-mode ansatz reads
\begin{align}
\label{4modes}
    \Psi=(a_1\varphi_8+a_2\varphi_9+b_1 \varphi_2+b_2 \varphi_3)e^{-i \mu_0 t}
\end{align}
where $\mu_0=\frac{1}{4}(\tmu_2+\tmu_3+\tmu_8+\tmu_9)$.  Now the phase differences $\phi_{a,b}$ and the population imbalances withing each of pairs of the modes $z_{a,b}$ are defined as above [see (\ref{za})] but with $\cN=|a_1|^2+|a_2|^2+|b_1|^2+|b_2|^2$. We also introduce the imbalance between populations of the pairs 
\begin{align}
z=\frac{1}{\cN}(|b_1|^2+|b_2|^2-|a_1|^2-|a_2|^2),    
\end{align}
as well as the phase mismatch 
\begin{align}
\phi=
\frac{1}{2}[\arg (a_2)+\arg (a_1)-\arg (b_2)-\arg (b_1)].
\end{align} 
In these variables the equations for the four-mode dynamics are governed by the Hamiltonian $H=H_a+H_b+H_{\rm in}$ where 
\begin{align}
\label{Ha}
H_a=&\sqrt{z_-^2-z_a^2}\cos\phi_a -\frac{\Lambda_a}{2}(z_a^2 {+}z_-^2),
\\
\label{Hb}
H_b=&{\epsilon} \sqrt{z_+^2-z_b^2}\cos\phi_b -\frac{\epsilon \Lambda_b }{2}(z_b^2 {+}z_+^2),
\end{align}
 $z_\pm=(1\pm z)/2$ describe population dynamics of the $a-$ and $b-$modes independently, while 
\begin{align}
\label{Hint}
     H_{\rm in}=
     &2\omega z+ \frac w2 (z^2-4z_a z_{{b}})
     \nonumber  \\
     &-\sqrt{(z_-+z_a)(z_++z_b)}[w_a(z_-+z_a)
     \nonumber \\
    &+w_b (z_++z_b)]\cos[(\phi_b-\phi_a)/2+\phi]
    \nonumber  \\
    &-\sqrt{(z_--z_a)(z_+-z_b)}[w_a(z_--z_a)
    \nonumber \\
    &+w_b (z_+-z_b)]\cos[(\phi_b-\phi_a)/{2}-\phi]
 \end{align}
with 
\begin{align*}
\omega=\frac{\tmu_2+\tmu_3- 2 \mu_0}{2(\tmu_9-\tmu_8)}=\frac{2\mu_0-\tmu_8-\tmu_9}{2(\tmu_9-\tmu_8)},
\\
    w_a=\frac{g \cN}{{\tmu_9-\tmu_8}}\int \varphi_2^3 \varphi_{8} d x, \quad w_b=\frac{g\cN}{{\tmu_9-\tmu_8}}\int \varphi_{8}^3 \varphi_2 d x,
    \\
    w=\frac{g\cN}{{\tmu_9-\tmu_8}}\int \varphi_2^2\varphi_{8}^2 d x,
\end{align*}
describes interaction between $a-$ and $b-$modes originated by the nonlinearity. 

The equations of motion are obtained as Hamiltonian equations
\begin{align}
     \frac{d z_{a,b}}{d\tau}=\pdv{H}{\phi_{a,b}},\quad \frac{d\phi_{a,b}}{d\tau}=-\pdv{H}{z_{a,b}}, 
 \\ 
   \frac{dz}{d\tau}=\pdv{H}{\phi}, \quad \frac{d\phi}{d\tau}=-\pdv{H}{z}.
\end{align}
Obviously, at zero nonlinear hopping, $w_{a,b}=w=0$, we have $H_{\rm in}=2\omega z$. Then $\phi$ is a cyclic variable and thus $z$ is constant. In this case we recover the two-mode dynamics described previously by the  equations (\ref{eq:z_phi}) and (\ref{eq:z_phi_b}) in the form  
 %\begin{subequations}\label{eq:4m_dynamics}
\begin{align}
\label{eq:zab}
     \frac{dz_a}{d\t}=- \sqrt{z_-^2- z_a^2} \sin \phi_a, \,\,\,  \frac{d\phi_a}{d\t}=\Lambda_a z +\frac{ z_a \cos \phi_a}{\sqrt{z_-^2-z_a^2}},\\
\label{eq:z}
     \frac{dz_b}{d\t}=- \epsilon \sqrt{z_+^2- z_b^2} \sin \phi_b, \,\,\,  \frac{d\phi_b}{d\t}=\epsilon\Lambda_b z +\frac{ \epsilon z_b \cos \phi_b}{\sqrt{z_+^2-z_a^2}},
 \end{align}
%\end{subequations}
accounting for the fact that the total number of atoms $\cN$ can be distributed differently between the pairs of $a-$ and $b-$modes as determined by $z_-$ and $z_+$ respectively. The fixed points of Eqs. (\ref{eq:zab}) and (\ref{eq:z}) are given by:
\begin{align}
\label{za-four-gminus}
        z_a^\pm=\pm \sqrt{z_-^2 - {1}/{\Lambda_a^2}}, \quad  \phi_a=0,
        \\
        \label{zb-four}
    z_b^\pm=\pm \sqrt{z_+^2 - {1}/{\Lambda_b^2}}, \quad \phi_b=0,
\end{align}
for $g=-1$, and by 
\begin{align}
\label{za-four-gplus}
        z_a^\pm=\pm \sqrt{z_-^2 - {1}/{\Lambda_a^2}}, \quad  \phi_a^\pm=\pm \pi,
        \\
        \label{zb-four}
    z_b^\pm=\pm \sqrt{z_+^2 - {1}/{\Lambda_b^2}}, \quad \phi_b^\pm=\pm \pi,
\end{align}
for $g=1$.
The diversity of the dynamical regimes is now much  richer than that of the two-mode system described above. Below we concentrate only on some of them allowing for better understanding of the limitations of the two mode-model. 
 
\subsection{Switching}
\label{sec:switch} 
\begin{figure}%[h]
     \centering
     \includegraphics[width=0.49\linewidth]{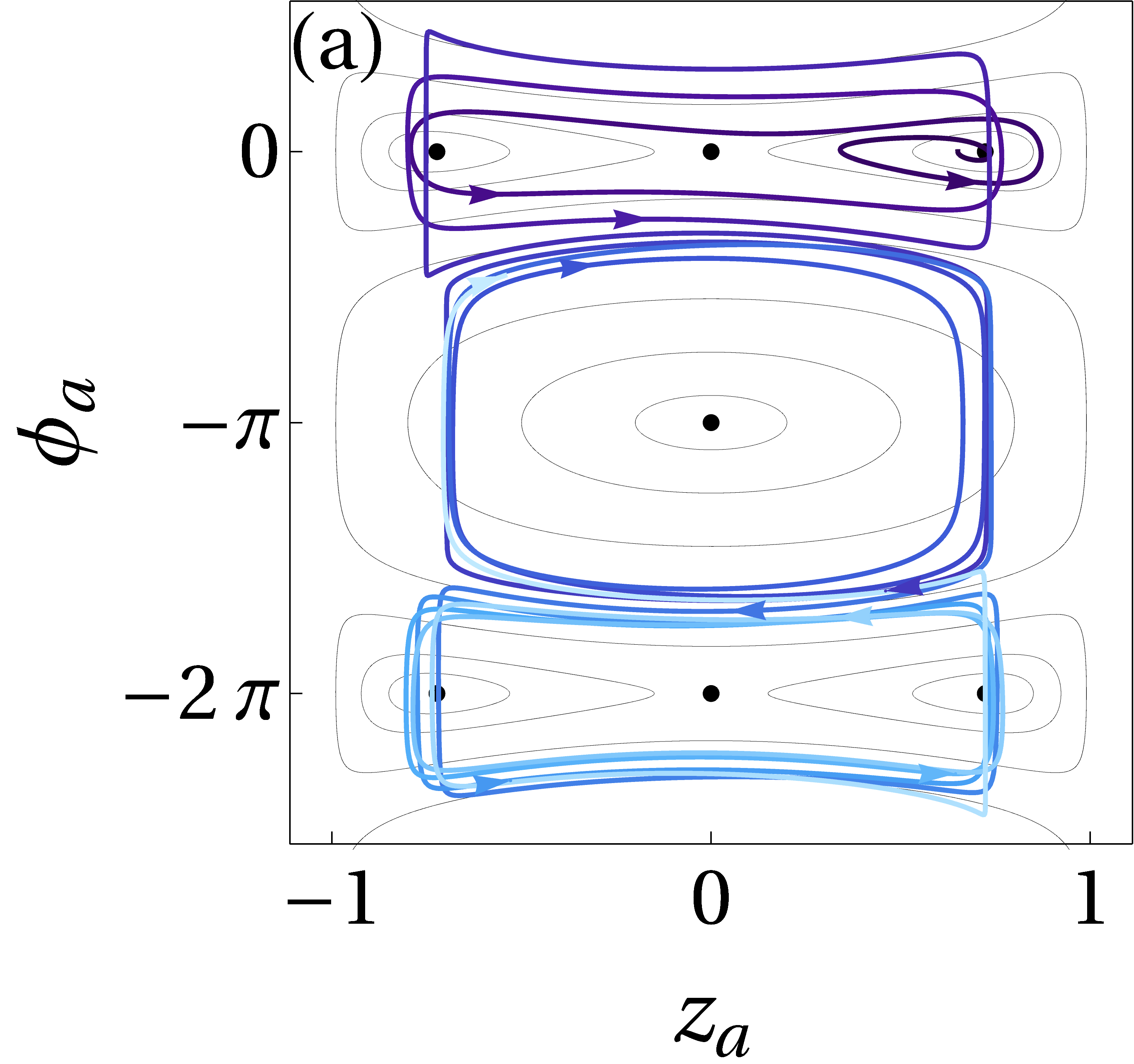}
      \includegraphics[width=0.49\linewidth]{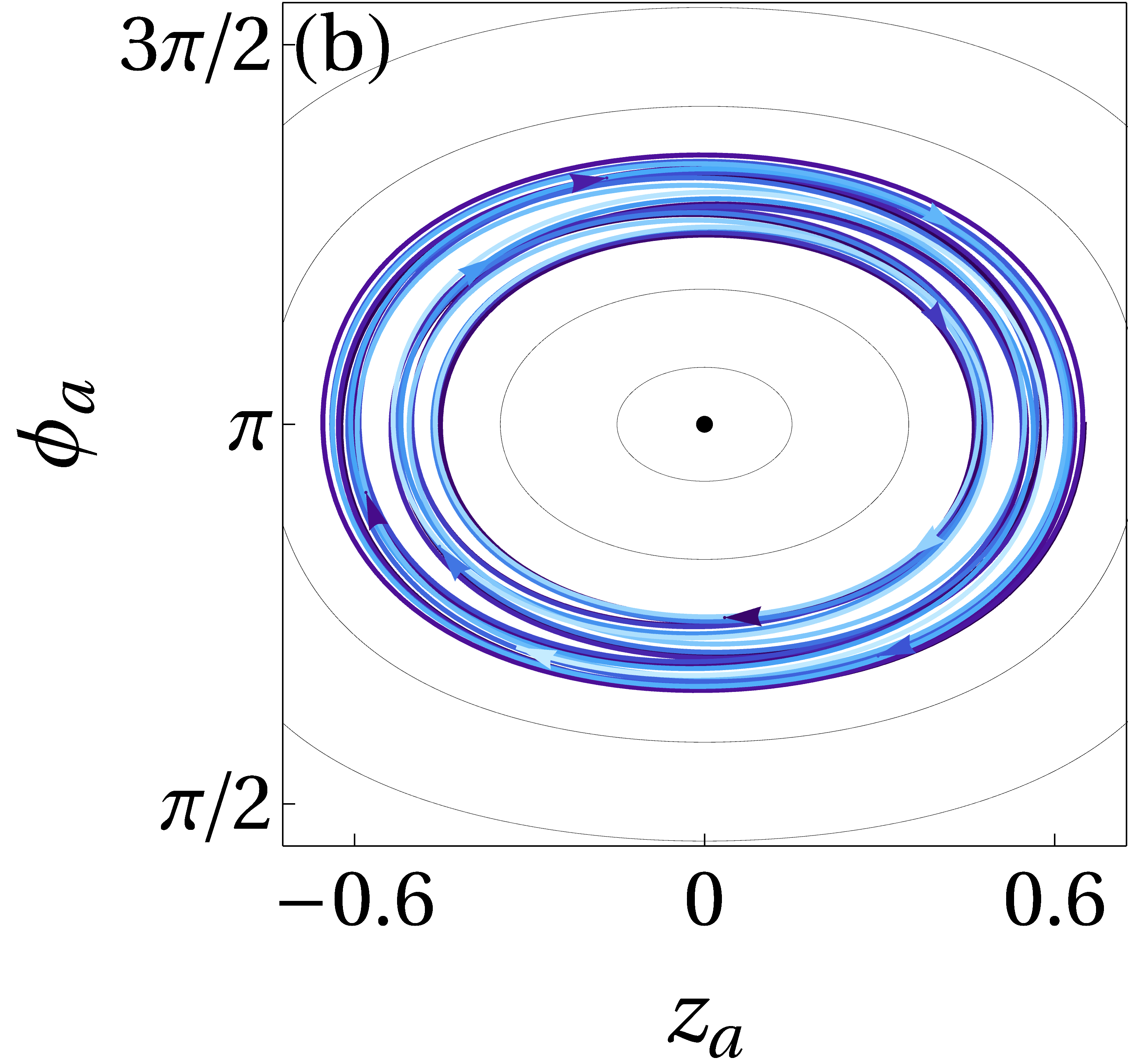}
     \caption{Projection of a phase trajectory of the four-mode approximation (\ref{eq:zab}), (\ref{eq:z}) on the $(z_a,\phi_a)$ plane for ${\Lambda_a=-1.46}$. The trajectory corresponds to the initial imbalance $z=-0.99$ ($0.5\%$ of the population belongs to the $b$-modes). From darker to lighter blue colours correspond to different types of the dynamics from earlier to later times (see the text). Light grey lines are the phase portrait at $z=-1$ (all atoms belong to the $a-$modes). Panels (a) and (b) show the projections corresponding to the orbits I and II of the underlying two-mode model in Fig.~\ref{fig:GPE_sol_g_neg}(a). The arrows mark the direction of evolution.}
     \label{fig:four_proj}
 \end{figure}

For numerical simulations, the nonlinear coefficients acquire the following values {$w_a\approx \pm 5.81\, \cN$, $w_b\approx \mp 9.14\,  \cN$ and $w\approx \pm 1.23\,\cN$ ($g=\pm 1$).} We start with the four-mode generalisation of the switching described in Fig.~\ref{fig:GPE_sol_g_neg}(a) and consider the dynamical system (\ref{eq:zab}), (\ref{eq:z}) with the initial condition in which 99.5\% of atoms being initially in the $a-$modes. Since the respective phase space is six-dimensional, to perform comparison with the counterpart two-mode model (\ref{eq:z_phi}), in Fig.~\ref{fig:four_proj} we show the projections of two trajectories on the $(z_a,\phi_a)$-plane.
  The trajectory starting near the self-trapping point {$(z_a,\phi_a)=(z_a^+, 0)$}   is highly sensitive to the population of the modes. Owing to the atoms transfer from $a-$ to $b-$modes the projection in Fig.~\ref{fig:four_proj}(a) (dark blue) increases its distance from the self-trapping point, and at some instant it becomes alike  the oscillatory trajectory III in Fig.~\ref{fig:GPE_sol_g_neg}. At this instant the switching occurs. At later times there occurs another switching to the oscillations around the center $(z_a,\phi_a)=(0,-\pi)$ (blue line), which follow by the subsequent switching (shown by  light blue line).
  
\begin{figure}
     \centering
     \includegraphics[width=\linewidth]{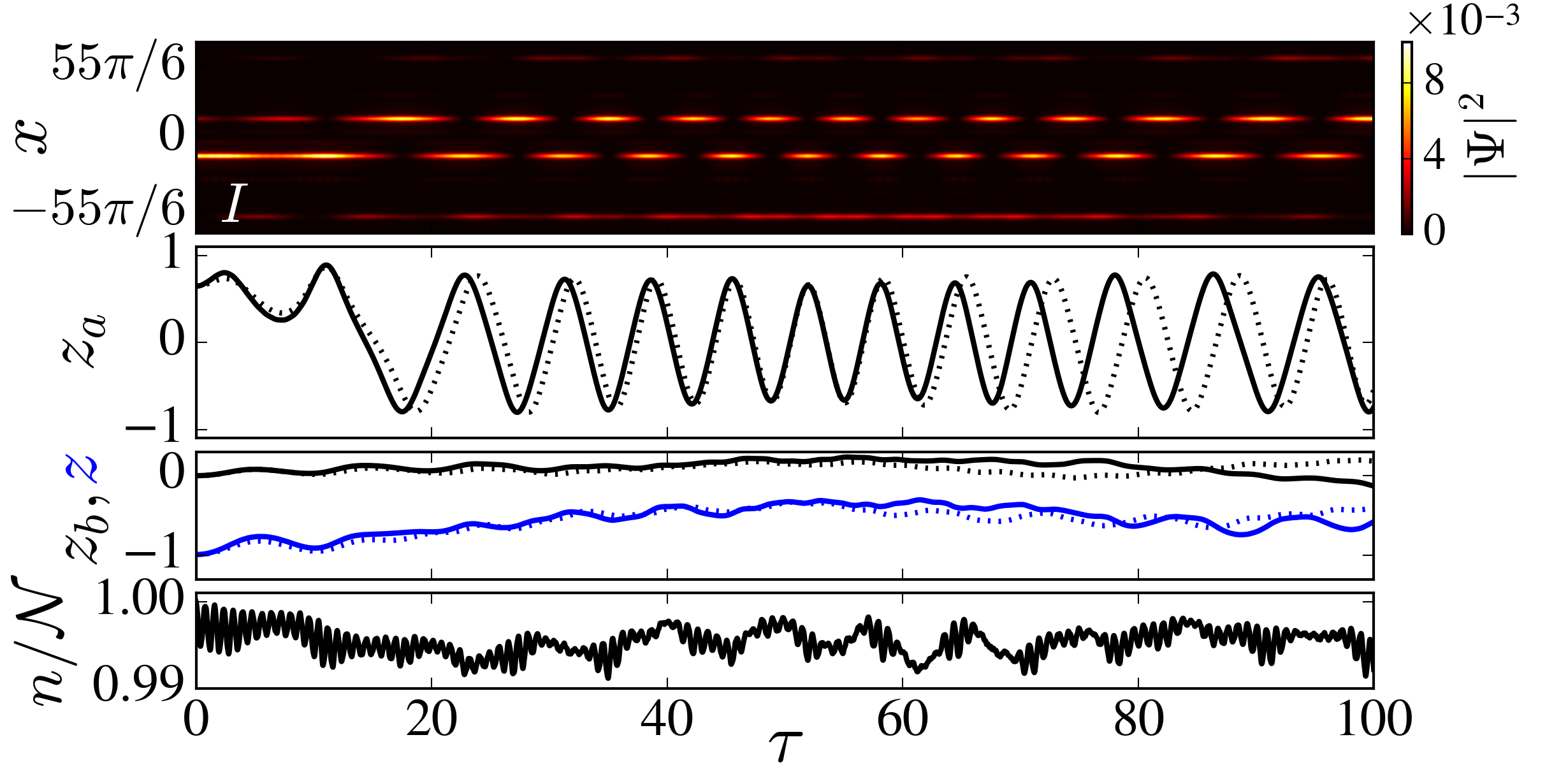}
     \caption{Numerical solution of the GPE~\eqref{eq:GP_nond} with the negative scattering length, that corresponds to switching dynamics shown in Fig.~\ref{fig:four_proj}(a) for the initial data $z=-0.99$, $z_a=0.9z_a^+\approx0.65$, $z_b=0$, $\phi_a=\phi_b=\phi=0$, with $\cN=1.57\times 10^{-2}$. The upper panel shows the density distributions, the middle panels show the population imbalances $z_a$, $z_b$ and $z$ for the numerical solution (solid lines) and the four-mode model (dotted lines), and the lower panel shows the fraction of population on the four modes initially excited $n/\cN$ {obtained from the projections on the a- and b-modes, analogously to \eqref{eq:n}.}}
     %$n=\int (\vphi_1^*+\vphi_2^*+\vphi_7^*+\vphi_8^*)\Psi  dx$}. }
     \label{fig:switching}
 \end{figure}

  The described evolution  [corresponding to the switching between the trajectories I, II and III in Fig.~\ref{fig:pportrait} (a)] is confirmed now with initial simultaneous excitation of $a-$ and $b-$modes, as shown in Fig.~\ref{fig:switching}. In the upper panel one can see a weak $b-$mode. The switching between self-trapping solutions and coherent oscillations occurs at $\tau\approx 15$. [Notice that the comparison should be considered qualitative, because the initial conditions for the dynamical systems (\ref{eq:zab}), (\ref{eq:z}) and (\ref{eq:z_phi}) are not exactly the same].
  
\subsection{Beatings}

Another phenomenon, that was not captured by the two-mode model, but can be observed  in the four-mode approximation (\ref{eq:zab}), (\ref{eq:z}), is beatings. In the trajectory shown in Fig.~\ref{fig:four_proj}(b) we observe a slow-time dynamics of changing the amplitude of the trajectory imposed on the fast rotations around the center. Obviously this the effect of slow variations of $z$. Indeed, it follows from (\ref{eq:z_phi_b}) and  the definition of the Hamiltonian, that  $dz/d\tau\propto w_{a,b},w$. In other words, such beatings are induced by the atomic exchange between the modes.  

\begin{figure}
     \centering
     \includegraphics[width=\linewidth]{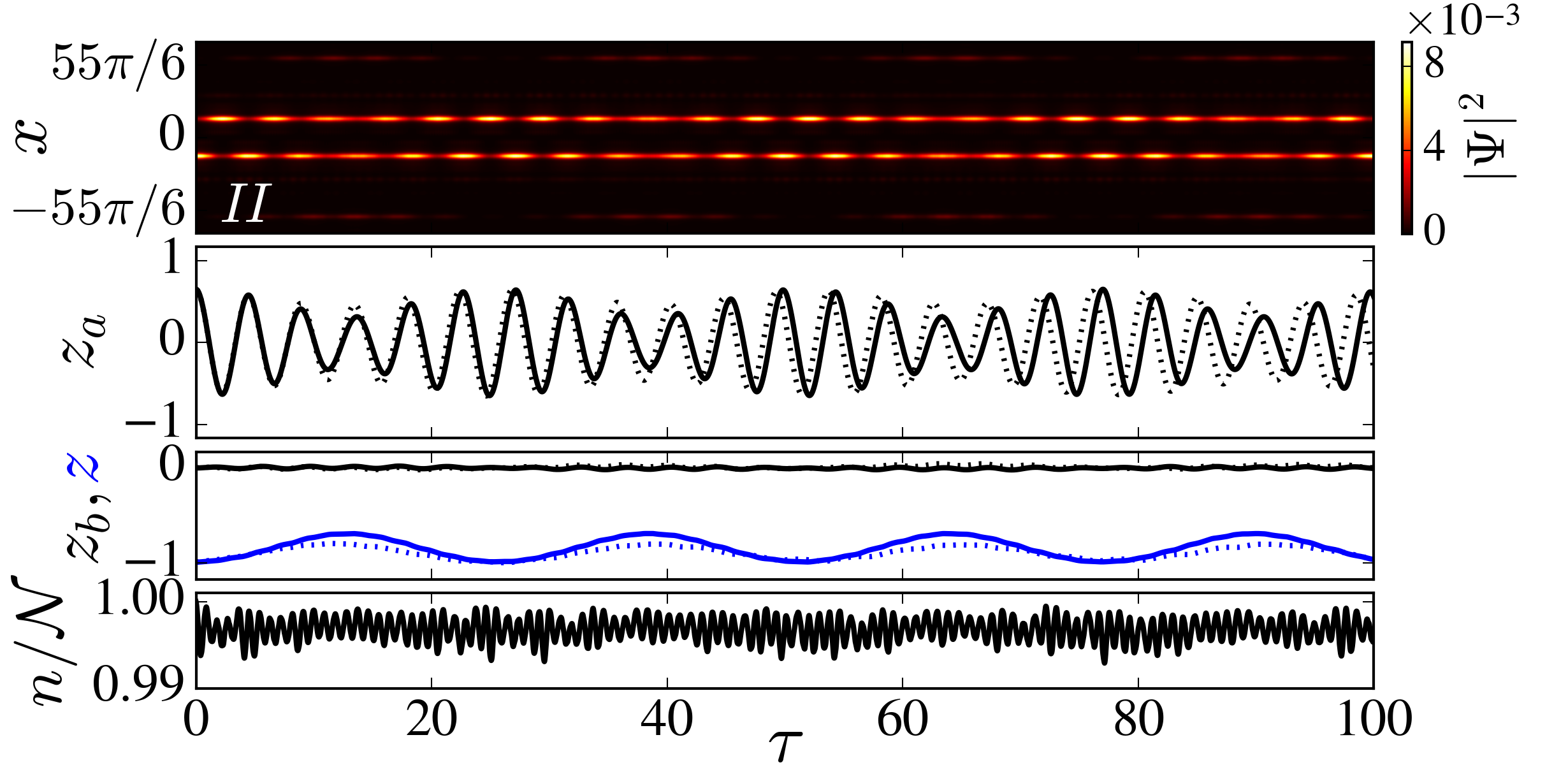}
     \caption{The same as in Fig.~\ref{fig:switching} but for the initial conditions $z=-0.99$, $z_a=0.9z_a^+\approx0.65$, $z_b=0$, $\phi_a=\pi$, $\phi=\phi_b=0$ corresponding to the trajectory in Fig.~\ref{fig:four_proj}(b).}
     \label{fig:beatings}
 \end{figure}

In Fig.~\ref{fig:beatings} we show the numerical comparison of the direct numerical simulations of the GPE, with four modes initially excited (the upper panel) with the dynamics governed by the four-mode approximation. Comparing the solid and dashed lines in the middle panels we conclude that beatings are indeed accurately predicted by the four-mode model.

\subsection{On weakly coupled self-trapped states}
 
\begin{figure}
    \centering
    \includegraphics[width=\linewidth]{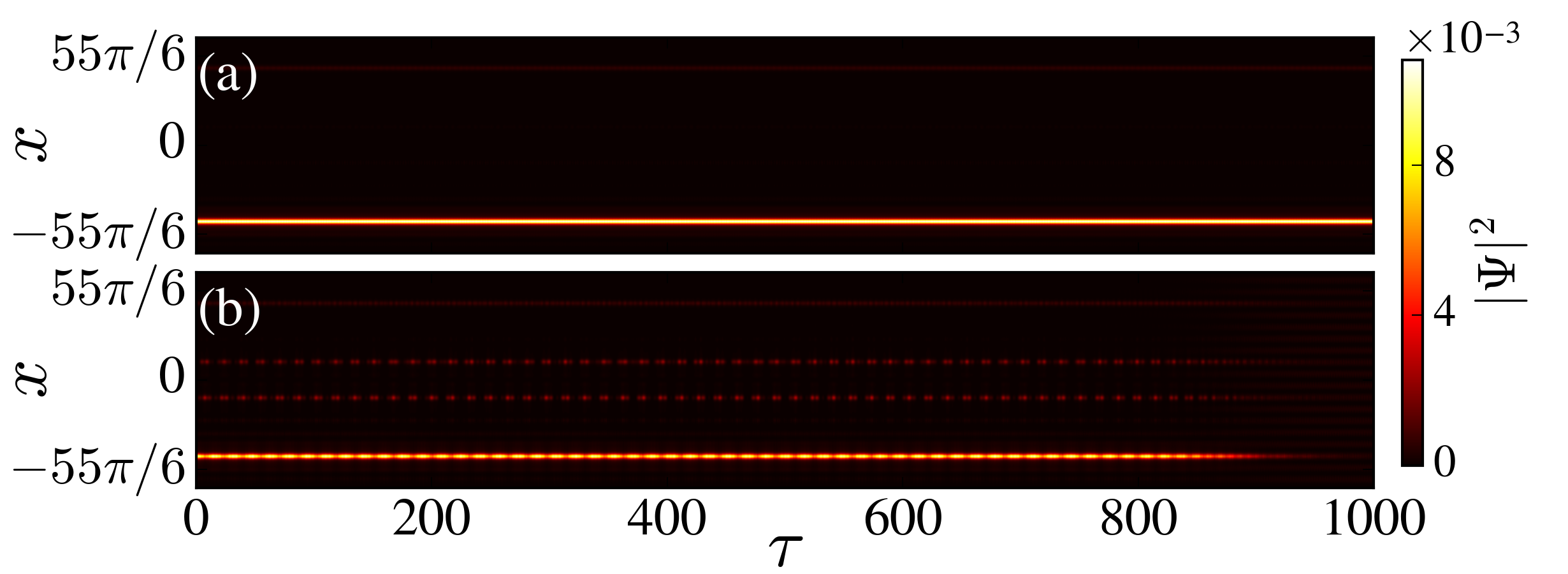}
    \caption{Numerical solutions of the GPE \eqref{eq:GP_nond} corresponding to the dynamics of the $b-$modes near self-trapping points for negative (a) and positive (b) scattering lengths, $\cN=1.57\times 10^{-2}$, with the initial data $z=0.99$ (95.5\% of all atoms initially in the $b-$modes), $z_a=0$, $z_b= 0.9 z_b^+ \approx 0.895 $, $\phi_a=\phi=0$ and (a) $\phi_b=0$ (b) $\phi_b=\pi$. }
    \label{fig:soliton}
\end{figure}

An essential difference between $a-$ and $b-$modes consists in much weaker linear coupling of the latter ones: $|\tmu_3-\tmu_2|\ll|\tmu_9-\tmu_8|$. This occurs due to greater spatial separation of the $b-$modes. Thus, if only one of the $b-$states, $\varphi_2$ or  $\varphi_3$, is excited, one can expect that even a very weak nonlinearity will transform it in a stable soliton. This is indeed what happens in the direct numerical simulations illustrated in Fig.~\ref{fig:soliton}. For a negative scattering length, we observe a stable soliton, whose norm undergoes weak oscillations (about 1\%) caused by the atom exchange with small amplitude $a-$modes also excited at $\tau=0$ [Fig.~\ref{fig:soliton}(a)]. This behaviour strongly differs from the switching that we observed for the $a$-modes near the self-trapping limit [cf. Fig.~\ref{fig:GPE_sol_g_neg}I].  Repulsive two-body interactions result in long living strongly asymmetric $b$-modes, but after some time $\tau\approx 900$ the nonlinearity results in the dispersion of the wave packet [Fig.~\ref{fig:soliton}(b)]. 
% \begin{figure}%[h]
%      \centering
%      %\includegraphics[width=\linewidth]{Figures/GPE_I_z2_-0.99.png}
%      %\includegraphics[width=\linewidth]{Figures/GPE_I_z2_-0.9.png}
%      \includegraphics[width=\linewidth]{Figures/GPE_I_z2_0.99.png}
%      %\includegraphics[width=\linewidth]{Figures/GPE_I_z2_0.99.png}
%      \caption{ The same as in Fig.~\ref{fig:switching} but for the initial conditions $z=0.99$ (95.5\% of all atoms are initially in $b-$modes). 
%      %\vk{In this case we can show all $z$ in the same panel. Also we have to define in the caption all lines (blue, black, dashed etc.)}
%      }
%      \label{fig:soliton}
%  \end{figure}

\section{  On preparation of initial states}

\label{sec:protocol}
 
 Now we briefly address the issue of exciting the desired atomic states, in particular those used in the present work. While this task may look challenging in view of the multitude
 %multiplicity 
of the states having very close energies, from the theoretical perspective the excitation of such modes is greatly facilitated by their nearly homogeneous spatial distribution (see Sec.~\ref{sec:linear}), which implies that the states with close energies are well separated in space.
%makes the state. Indeed, the task is greatly facilitated by nearly homogeneous spatial distribution of the modes making them well separated. 
Indeed, a two hump mode, with humps centered at $X_\pm$ and considered `separately' can be expanded over the complete set of the Gauss-Hermite functions (i.e., harmonic oscillator eigenstates) 
 $\xi_n(x)=e^{-x^2/(2\sigma^2)} \mathcal{H}_n(x/\sigma)$, where $\mathcal{H}_n$ is the $n$th Hermite polynomial:
%  \begin{equation}\label{eq:IC_gauss}
%      \Psi(x, 0)=\sqrt{\frac{\mathcal{N}}{4\pi}}\left(\displaystyle e^{- (x-X_-)^2/2}\pm e^{-(x-X_+)^2/2}\right),
%  \end{equation}
\begin{equation}
\label{eq:IC_gauss}
     \Psi(x, 0)=\sum_{n=0}^\infty c_n \left[ \xi_n(x-X_-) \pm \xi_n(x-X_+)\right].
 \end{equation}
In Eq. (\ref{eq:IC_gauss}), $c_n$ are projections of the mode over the states $\xi_n(x-X_\pm)$ and $\sigma>0$ is an optimisation parameter. Due to strong localization of the modes, one can find $\sigma$ such that only a few terms in (\ref{eq:IC_gauss}) are essential, i.e., representing, say, above $90\%$ of atoms of the mode.  Since however, the linear states have a multi-peak structure, i.e., in a pure linear system two Gauss-Hermite modes cannot generate  the desired  linear state, even if the initial harmonic-oscillator states are created in the correct locations $X_\pm$ before the quasi-periodic potential is switched on. This, difficulty, however can be overcome by using weak nonlinearity, which in the process of evolution makes the job of creating a nonlinear stationary state (belonging to a required family described in Sec.~\ref{sec:bifurcations}). After the weakly nonlinear mode  is formed, one can switch off the nonlinearity adiabatically, thus driving the system to the linear state, from which the respective family of the nonlinear modes bifurcates.

\begin{figure}
    \centering
    \includegraphics[width=\linewidth]{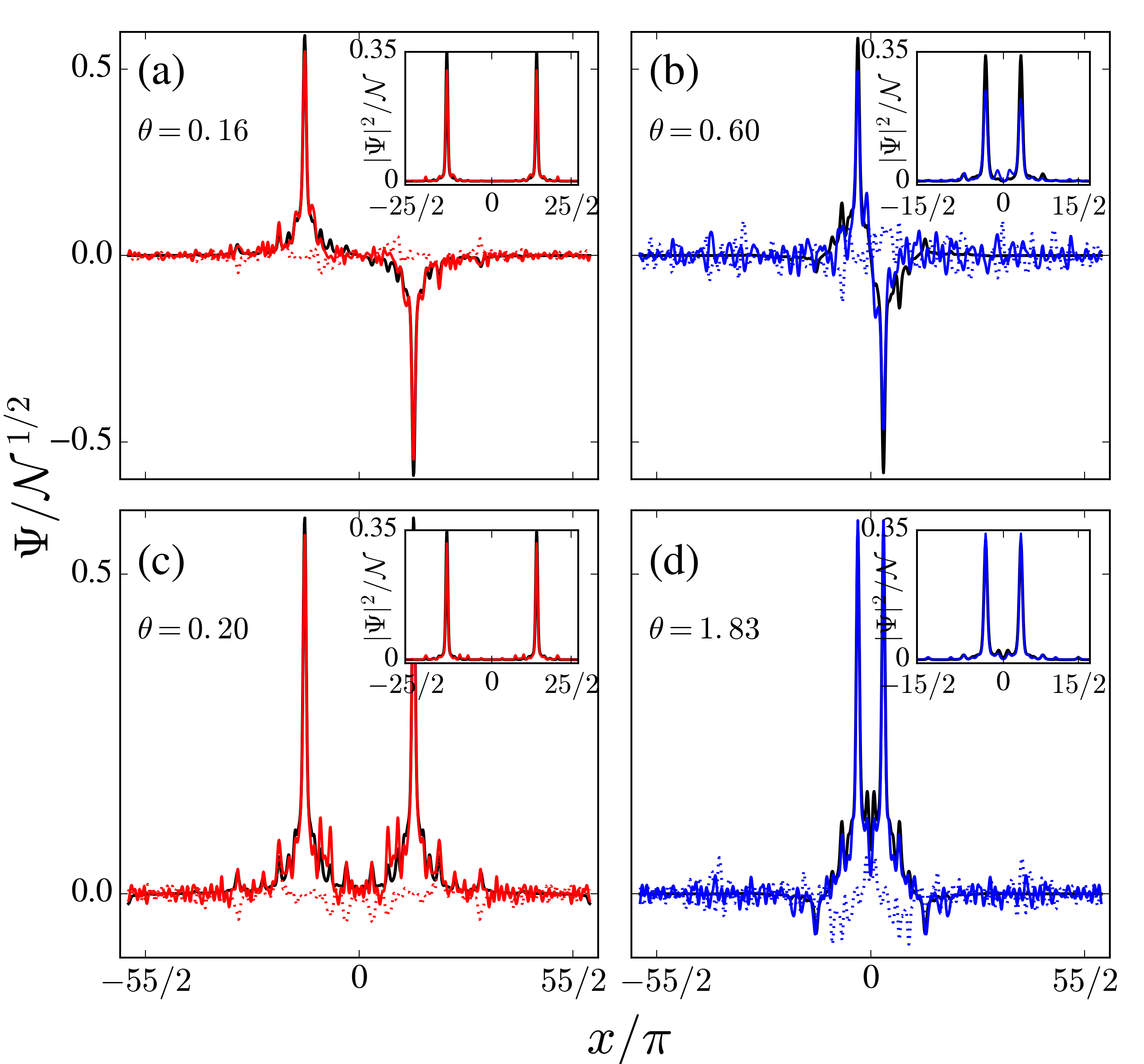}
    \caption{Numerical solutions of \eqref{eq:GP_nond} with an initial conditions \eqref{eq:IC_gauss}, evolved according to the procedure described in the text. Since such evolution for different cases ends up different phases $\theta$ (see the text) they are indicated in each panel. The black lines are the modes $\tpsi_{2,3,8,9}$ shown in Fig.~\ref{fig:two}. The colored solid and dotted lines represent distributions $\Re \Psi(\tau_n)$ and $\Im \Psi(\tau_n)$, respectively. The insets show the distributions of the atomic densities. In all panels $\sigma=1.3$.}
    \label{fig:Gaussian}
\end{figure}

In Fig.~\ref{fig:Gaussian} we illustrate direct numerical implementation of the described  procedure. 
 At $\tau=0$ for the even ("$+$") and odd ("$-$") modes we consider  the modes considered $n=0,4$ for the modes $\tpsi_{2,3,8}$ and $n=0,1$ for the mode $\tpsi_9$. Such initial condition evolves in a BEC with a negative scattering length, were we have chosen the attractive nonlinearity to be  $\cN=10^{-3}$ until the instant $\tau=10$, after which we start to adiabatically decrease the non-linearity according to the law $ 
    \cN(\tau)=\cN(0)(1-\alpha \tau)$ 
where we have used $\alpha=0.26$ (recall that $\tau$ is the slow time), until the linear regime is reached. During such evolution the order parameter $\Psi(x,\tau)$ may acquire additional, generally speaking inhomogeneous, phase. In the panels of Fig.~\ref{fig:Gaussian} we indicate the values $\theta$ of such phases computed in the points of maxima of atomic densities, i.e., at $x$ where $|\Psi(x,\tau_{\rm fin})|^2$ with
$\tau_{\rm fin}=10+1/\alpha$, are maximal.

Finally, we notice that the proposed approach for preparing the desired state is not the only one, as soon as the explored linear modes are Bloch states in the center of the Brillouin zone of the respective superlattice approximation. Meantime, further improvement of the method can be achieved by using additional optimisation parameters, in particular, phase distributions resulting in $x-$independent phase at $\tau_{\rm fin}$. This task, however goes beyond the scope of this work.
 
 \section{Conclusion}
 
 A BEC in a quasi-periodic potential is a versatile system allowing for the observation of several phenomena which are characterized by nonlinear dynamics of two or more coupled BECs. In order to emulate a bosonic Josephson junction, we considered a one-dimensional condensate in a bi-chromatic optical lattice emulating bosonic Josephson junction, resembling the setting with a double-well trap, in both positive and negative scattering lengths. Approximating the almost periodic potential by a periodic one, we have shown that the system possesses a memory effect when each subsequent more accurate approximation inherits the information from all previous rational approximations both in  spectral and real-space distribution of the modes. Below the mobility edge, we identified different pairs of modes (even and odd ones) allowing for constructing strongly localized linearly coupled states, alike those known for a double-well trap, but localized due to the interference effect rather than due to confining potential wells. Studying the dynamics of such couples of modes we obtained coherent oscillations, which in our case display such additional phenomena as switching between seftrapping and different oscillatory regimes. These phenomena are related to the excitation of other localized modes due to the nonlinearity, and can be captured by the four-mode model, which was also deduced. We also discussed bifurcations of the nonlinear stationary modes and obtained minigap solitons being a peculiarity of a quasi-periodic system.
 
 The phenomena described here can also be observed in other physical systems, like for example light propagation in a quasi-periodic Kerr medium, and most likely allows for generalization for a two-dimensional setting, which makes the problem to be of a fairly general interest for physical applications beyond the theory of matter waves.

\begin{acknowledgments}
H.C.P. and V.V.K. acknowledge financial support from the Portuguese Foundation for Science and Technology (FCT) under Contracts PTDC/FIS-OUT/3882/2020 and UIDB/00618/2020.
 The work of D.A.Z. was supported by   Priority 2030 Federal Academic Leadership Program. 
 
\end{acknowledgments}

\end{document}